\documentclass[12pt]{article}
\usepackage[dvips]{graphicx}
\def\Tr{\mathrm{Tr}}
\def\min{\mathrm{min}}

\def\x{\mathbf{x}}
\def\H{\widehat{H}}
\usepackage{amsmath}
\usepackage{amssymb}
\def\Tr{\mathrm{Tr}}
\def\x{\mathbf{x}}

\begin{document}

\centerline{ \Large \bf  Replica Symmetry Breaking Condition
Exposed} \centerline{ \Large \bf by Random Matrix Calculation of
Landscape Complexity}

\vskip 0.4cm

\centerline{ \bf Yan V. Fyodorov and Ian Williams}

\vskip 0.3cm

\centerline{School of Mathematical Sciences, University of
Nottingham, Nottingham NG72RD, England}

\vskip 0.3cm

\begin{abstract}
We start with a rather detailed, general discussion of recent
results of the replica approach to statistical mechanics of a
single classical particle placed in a random $N (\gg
1)$-dimensional Gaussian landscape and confined by a spherically
symmetric potential suitably growing at infinity. Then we employ
random matrix methods to calculate the density of stationary
points, as well as minima, of the associated energy surface. This
is used to show that for a generic smooth, concave confining
potentials the condition of the zero-temperature replica symmetry
breaking coincides with one signalling that {\it both} mean total
number of stationary points in the energy landscape, {\it and} the
mean number of minima are exponential in $N$. For such systems the
(annealed) complexity of minima vanishes cubically when
approaching the critical confinement, whereas the cumulative
annealed complexity vanishes quadratically. Different behaviour
reported in our earlier short communication [Y.V. Fyodorov et al.,
{\it JETP Lett.} {\bf 85}, 261, (2007)] was due to non-analyticity
of the hard-wall confinement potential. Finally, for the simplest
case of parabolic confinement we investigate how the complexity
depends on the index of stationary points. In particular, we show
that in the vicinity of critical confinement the saddle-points
with a positive annealed complexity must be close to minima, as
they must have a vanishing {\it fraction} of negative eigenvalues
in the Hessian.

\end{abstract}

\section{Introduction and General Discussion}

After more than three decades of intensive research, understanding
statistical mechanics of disordered systems still remains a
considerable challenge to both theoretical and mathematical
physics community. In the situations where an interplay between
thermal fluctuations and those due to quenched disorder is
essential, one of the central problems is to find the averaged
value of the free energy of the system as a function of
temperature $T$. Namely, given an energy function which assigns a
random value ${\cal H}(\bf{x})$ to every point ${\bf x}\in \Sigma$
of a given configuration space $\Sigma$, the task is to calculate
\begin{equation}\label{freendef}
F=-T\,\left\langle\ln{Z_\beta}\right\rangle,\quad
Z_{\beta}=\int_{\Sigma} \exp{-\beta{\cal H}({\bf x})}\, d {\bf
x}\,.
\end{equation} Here and henceforth the angular
brackets stand for the disorder averaging, $\beta=1/T$ is the
inverse temperature, and $Z_{\beta}$ is known as the partition
function. In the case when the configuration space consists of
discrete points (e.g. the hypercube $\Sigma=\{-1,1\}^N$) the
integration is replaced by an appropriate summation.

 Confronted with the problem of performing such a calculation explicitly,
theoretical physicists almost invariably choose to employ the
so-called replica trick - a powerful, yet heuristic way of
extracting the above average from the positive integer moments of
the partition function. The trick amounts to exploitation of the
formal identity
\begin{equation}\label{replica}
\left\langle\ln{Z_{\beta}}\right\rangle=\lim_{n\to
0}\frac{1}{n}\ln{\left\langle Z_{\beta}^n\right\rangle},\quad
Z^n=\int_{\mathbb{R}^N} e^{-\beta\sum_{a=1}^n{\cal H}({\bf
x_{a}})}\prod_{a=1}^n d {\bf x}_{a}\,.
\end{equation}
Indeed, assuming the distribution of random energies ${\cal
H}({\bf x})$ to be Gaussian with the prescribed covariance in the
configuration space, the moments can be readily calculated for all
values $n=1,2,...$. Unfortunately, for all systems of interest
those moments grow too fast with $n$ to allow a unique restoration
of the probability distribution of the partition function, hence
prohibiting the honest evaluation of the expectation value of the
logarithm. Yet, physicists managed to employ their fine intuition
and developed a refined and efficient heuristic machinery of
performing a mathematically ill-defined continuation $n\to 0$. In
this way they were able to reconstruct highly nontrivial patterns
of behaviour typical for disordered systems, the most intricate
scenario being based on the notion of {\it Replica Symmetry
Breaking} (RSB). According to that scenario first developed by
Parisi in the context of the so-called mean-field models of spin
glasses, see \cite{MPV} as the standard reference and
\cite{Parisinew} for a recent account, below a certain transition
temperature (frequently referred to as the de-Almeida-Thouless
\cite{AT} temperature) the Boltzmann-Gibbs measure on the
configuration space is effectively decomposed into a multitude of
the so-called "equilibrium states" (this phenomenon is frequently
referred to as an {\it ergodicity breaking}). The number of those
states is argued to grow exponentially with the number of
effective degrees of freedom, and they are believed to be
organized in the configuration space in an intricate hierarchical
way. Essentially, the structure is thought to represent a picture
of valleys within valleys within valleys, etc., in an effective
random (free) energy landscape\cite{TAP} arising in such models.
As that unusual picture arouse via the application of a somewhat
problematic replica trick, it is important to mention that a few
rather nontrivial results emerging in the framework of the Parisi
scenario were recently recovered by rigorous mathematical methods.
These important developments are mainly due to recent seminal
works by Talagrand \cite{Talagrand1,Talagrand}, based on earlier
results by Guerra \cite{Guerra}, see also interesting works by
Aizenman, Sims and Starr \cite{ASS}. In particular, Talagrand was
able to demonstrate that the equilibrium free energy emerging
naturally in the Parisi scheme of RSB is indeed the correct
thermodynamic limit of the free energy, both for the paradigmatic
Sherrington-Kirkpatrick model \cite{Talagrand}, and for the
so-called spherical model of spin glasses\cite{Talagrand1}.

One of the simplest, yet nontrivial representatives of a
disordered system is arguably a single classical particle placed
in a superposition of a random Gaussian $V({\bf x})$ and a
deterministic confining potential $V_{con}({\bf x})$, with ${\bf
x}\in\Sigma= \mathbb{R}^N$. It turned out to be a surprisingly
rich model, characterised by a non-trivial dynamical behaviour as
well as interesting thermodynamics. Works by Mezard and Parisi
\cite{MP}, and Engel \cite{Engel} used the replica trick to
calculate the free energy of such a system with the simplest,
parabolic confinement $V_{con}({\bf x})=\frac{1}{2}\mu{\bf
x}^2\,,\mu>0$ and after further employing the so-called Gaussian
Variational Ansatz (GVA) revealed the existence of a
low-temperature phase with broken ergodicity. They were followed
by Franz and Mezard \cite{FM} and Cugliandolo and Le Doussal
\cite{longtime} papers on the corresponding dynamics revealing
long-time relaxation, aging, and other effects typical for glassy
type of behaviour at low enough temperatures. The nature of the
low-temperature phase was found to be very essentially dependent
on the type of correlations in the random potential, specified via
the covariance function chosen in the form ensuring stationarity
and well-defined large-$N$ limit as
\begin{equation}\label{cov1}
\left\langle V\left({\bf x}_1\right) \, V\left({\bf
x}_2\right)\right\rangle=N\,f\left(\frac{1}{2N}({\bf x}_1-{\bf
x}_2)^2\right)\,,
\end{equation}
with brackets standing for the averaging over the Gaussian
potential distribution. Namely, if the covariance $f(x)$ decayed
to zero at large arguments, the description of the low temperature
phase was found to require only the so-called one-step replica
symmetry breaking (1RSB) Parisi pattern. In contrast, for the case
of long-ranged correlated potentials with $f(x)$ growing with $x$
as a power-law\footnote{To be more precise, at large separations
$\left\langle \left[V\left({\bf x}_1\right)- V\left({\bf
x}_2\right)\right]^2\right\rangle\propto \left({\bf x}_1-{\bf
x}_2\right)^{2\gamma}, \, 0<\gamma<1$. However one can easily
satisfy oneself that the difference is immaterial for the
free-energy calculations.} the full infinite-hierarchy Parisi
scheme of replica symmetry breaking (FRSB) had to be used instead.

Based on formal analogies with the Hartree-Fock method Mezard and
Parisi\cite{MP} argued that GVA-based calculations should become
exact in the limit of infinite spatial dimension $N$. In a recent
paper \cite{FS} the replicated problem was reconsidered in much
detail by an alternative method which directly exposed the degrees
of freedom relevant in the limit $N\to \infty$, and employed the
Laplace (a.k.a. saddle-point) evaluation of the integrals. The
method allowed also to perform calculations for any fixed sample
size $|{\bf x}|^2<L=R\sqrt{N}$. The effective radius $R<\infty$
can be used as an additional control parameter, and the results in
the limit $R\to \infty$ and fixed $\mu>0$ indeed reproduced those
obtained by GVA in \cite{MP,Engel}. In addition, the research in
\cite{FS} and the subsequent work \cite{FBprl} revealed the
existence of a non-trivial class of random potentials overlooked
in earlier papers\cite{MP,Engel},
 the simplest representative being the case of logarithmically growing
correlations $f(x)=f_0-g^2\ln{(x+a^2)}$, with $f_0,g,a$ being some
positive constants. Indeed, for such a choice (and its
generalizations\cite{FBprl}) the resulting free energy in the
thermodynamic limit $R\to \infty$ turns out to coincide precisely
with that appearing in the celebrated Generalized Random Energy
Model (GREM) by Derrida\cite{Derrida,GREM}, which is, in
particular, known to describe directed polymers on a tree with
disordered potentials\cite{DeSp}. Moreover, by comparing the
results of\cite{FBprl} with earlier renormalization-group analysis
of same system for finite dimensions $1\le N<\infty$, see
Carpentier and Le Doussal \cite{CL}, it was conjectured that
essentially the same type of behavior should survive in
finite-dimensional systems. Note, that understanding the behaviour
of the model for finite $N$ in full generality remains a
challenging problem, see e.g. \cite{TD} for a recent discussion.

According to a general wisdom, many peculiar features of systems
with quenched disorder as compared to their more ordered
counterparts are largely due to a multitude of nearly degenerate
(metastable) states making the associated (free) energy landscape
extremely corrugated, see e.g an insightful paper \cite{BBM}
trying to understand the gross structure of the energy landscapes
describing elastic manifolds in presence of a random pinning
potential. In particular, for the paradigmatic mean-field models
of spin glasses a considerable effort has gone into trying to
understand what changes the intricate scenario of the replica
symmetry breaking interpreted as ergodicity breaking (see our
earlier discussion of the Parisi scenario) implies in the
statistics of the associated free energy landscape (the so-called
Thouless-Anderson-Palmer (TAP)\cite{TAP} variational functional).
For an instructive review of the recent efforts in this direction
a reader may wish to look through \cite{Parisirev}, or to read
directly representative original works on this topic
\cite{metast}-\cite{metast5}.

For the model of interest for us in this paper - a single particle
in a random Gaussian potential - the necessity (and difficulty) of
adequately taking into account multiple minima in the energy
landscape is well-known since the early work of
Engel\cite{Engelold} who pointed out a non-perturbative character
of the problem. In fact, similar non-perturbative effects are
believed to invalidate the famous "dimensional reduction"
phenomenon in the random-field Ising model, see
\cite{Villain,Natt} and references therein. It is therefore
important to develop methods and tools of statistical
characterization of relevant features of $N-$dimensional random
energy landscapes. Ideally, one would like to obtain a detailed
knowledge on typical positions and depths of various minima and
their mutual correlations. In such a general formulation the
problem looks extremely challenging mathematically, although in
the dimension $N=1$ considerable progress can be achieved
\cite{1Db}.

Obviously, the first meaningful step in this direction consists of
counting the expected number of stationary points of different
types (minima, maxima and saddles of various indices). The
simplest, yet already a non-trivial problem of this sort is to
find the mean number $\langle\#_s\rangle$ of {\it all} stationary
points, irrespective of their index. Assuming that the landscape
is described by a sufficiently smooth random function ${\cal H}$
of $N$ real variables ${\bf x}=(x_1,...,x_N)$ the problem amounts
to finding all solutions of the simultaneous stationarity
conditions $\partial_k {\cal H}=0$ for all $k=1,...,N$, with
$\partial_k$ standing for the partial derivative
$\frac{\partial}{\partial x_k}$. The total number $\#_s(D)$ of its
stationary points in any spatial domain $D$ is then given by
$\#_s(D)=\int_D \rho_s({\bf x}) \, d{\bf x}$, with $\rho_s({\bf
x})$ being the corresponding density of the stationary points. The
ensemble-averaged value of such a density can be found according
to the multidimensional analogue of the so-called Kac-Rice
formula:
\begin{equation}\label{KR}
\langle\rho_{s}({\bf x})\rangle=\left\langle
|\det{\left(\partial^2_{k_1,k_2} {\cal H}\right)}|
\prod_{k=1}^N\delta(\partial_k {\cal H})\right\rangle,
\end{equation}
where $\delta(x)$ stands for the Dirac $\delta-$ function. For the
original Kac and Rice papers in $N=1$ see \cite{Kac,Rice}, the
multidimensional generalization can be found e.g. in \cite{KR} and
\cite{CPRW}, and \cite{AdlerT} contains further extensions and a
proof in general setting. Note importance of keeping the modulus
in (\ref{KR}), as omitting it would yield instead of the density
the object related to the Euler characteristics of the surface,
see \cite{Kurchan} and discussions in \cite{AdlerT} (see also the
related identity (\ref{ident}) below).

Similarly, if one is interested in counting only minima, the
corresponding mean density can be written as
\begin{equation}\label{KRM}
\langle\rho_{m}({\bf x})\rangle=\left\langle
\det{\left(\partial^2_{k_1,k_2} {\cal
H}\right)}\theta\left(\partial^2_{k_1,k_2} {\cal H}\right)
\prod_{k=1}^N\delta(\partial_k {\cal H})\right\rangle.
\end{equation}
Here and henceforth we use the notation $\theta(x)$ for the
Heaviside step-function, i.e. $\theta(x)=1$ for $x>0$, and zero
otherwise. The corresponding matrix $\theta-$factor in (\ref{KRM})
selects only stationary points with positive definite Hessians,
which are minima.

In the context of the problematic of glassy systems calculation of
those quantities for the simplest case $N=1$ were presented in
\cite{1Da}. Actually, low-dimensional cases $N=(1,2,3)$ were known
much earlier, starting from the classical papers by
Longuet-Higgings on specular light reflection from a rough sea
surface\cite{sea}, and the work on laser speckle patterns by
Weinrib and Halperin\cite{laser}. The case $N=3$ was also
considered in early work \cite{halp} in the context of behaviour
of a {\it quantum} particle in a random potential, and recently in
\cite{VM}. A useful summary of related efforts in various areas of
applied mathematics and probability theory can be found in a
recent book by Adler and Taylor\cite{AdlerT}.

 Random landscapes relevant in statistical physics are however
mainly high-dimensional. In particular, general interest in
landscapes behind glass-forming mixtures of various types led to a
few numerical investigations of the structure of their critical
points, see e.g. \cite{numglass}-\cite{numglass3}. Independently,
interest in counting extrema of some random high-dimensional
surfaces arose also in string theory where an effective string
landscape is believed to possess a huge number of possible minima,
each of which could describe a potential universe, see
\cite{string} for further discussion, references and actual
calculations. At the same time, no analytical results on numbers
of critical points were available for high-dimensional Gaussian
landscapes until very recently, before it was realized that
obtaining them requires exploitation of well-developed
random-matrix based techniques \cite{YFglass}. For the $N-$
dimensional Gaussian random surface with parabolic confinement,
i.e. for the function
\begin{equation}\label{Hpar}
{\cal H}_{par}({\bf x})=\frac{\mu}{2}\,{\bf x}^2+V({\bf x}),\quad
\mu>0
\end{equation}
the total expected number of stationary points calculated from the
Kac-Rice density (\ref{KR}) turned out to be given by
\cite{YFglass}:
\begin{equation}\label{tot}
\langle\#_s\rangle=\frac{1}{\mu^N}\left\langle|\det(\mu
\hat{I}_N+\hat{H})|\right\rangle\,,
\end{equation}
where $\hat{H}$ stands for $N\times N$ Hessian matrix of second
derivatives of the Gaussian part of the potential:
$H_{ij}\equiv\partial_{ij}^2 V({\bf x})\,,\,\{(i,j)=1,...,N\}$,
and $\hat{I}_N$ for the identity matrix, and we took into account
that the distribution of the Hessian is position-independent due
to translational invariance of the covariance function
(\ref{cov1}), see Eq.(\ref{JPD1}) below . We see that the problem
basically amounts to evaluating ensemble average of the absolute
value of the characteristic polynomial of a certain real symmetric
random matrix, whose entries are centered Gaussian variables due
to the nature of the underlying potential V({\bf x}). The form of
the covariance function Eq.(\ref{cov1}) implies after a simple
calculation the following covariances of the entries $H_{ij}\,$:
\begin{equation}\label{JPD1}
\left\langle H_{il}H_{jm}\right\rangle=\frac{f^{''}(0)}{N}\left[
\delta_{ij}\delta_{lm}+\delta_{im}\delta_{lj}+
\delta_{il}\delta_{jm}\right]\,.
\end{equation}
where here and henceforth in the paper the number of dashes
indicates the number of derivatives taken.
 This allows one to write down the density of the joint
probability distribution (JPD) of the matrix $H$ explicitly as
\cite{YFglass}
\begin{equation}\label{JPD2}
{\cal P}(\hat{H})d\hat{H}\propto
d\hat{H}\exp\left\{-\frac{N}{4f^{''}(0)}
\left[\mbox{Tr}\left(\hat{H}^2\right)-\frac{1}{N+2}
\left(\mbox{Tr} \hat{H}\right)^2\right]\right\}\,,
\end{equation}
where $d\hat{H}=\prod_{1\le i\le j\le N} dH_{ij}$ and the
proportionality constant can be easily found from the
normalisation condition. Although such a JPD is invariant with
respect to rotations $\hat{H}\to \hat{O}^{-1}\hat{H}\hat{O}$ by
orthogonal matrices $\hat{O}\in O(N)$ it is apparently different
from the standard one typical for the so-called Gaussian
Orthogonal Ensemble (GOE) of random symmetric matrices (the
standard introduction into the Random Matrix Theory is
\cite{Mehta}). A peculiarity of this ensemble is particularly
manifest via the following identity that holds for an {\it
arbitrary} fixed matrix $\hat{A}$ \cite{VM}:
\begin{equation}\label{ident}
\left\langle\det(\hat{A}+\hat{H})\right\rangle=\det{\hat{A}}\,.
\end{equation}
In particular, this identity gives another evidence for importance
of keeping the modulus in formulae like Eq.(\ref{tot}), where
omitting the modulus would trivially yield unity due to
(\ref{ident}). An elegant general verification of (\ref{ident})
directly from (\ref{JPD1}) can be found in \cite{VM}.
Alternatively, in Appendix A we demonstrate its validity for real
symmetric matrices by a specific method exploiting the very
meaning of the matrix $\hat{H}$ as the Hessian matrix, as the
latter is related to topological properties of the underlying
gradient field.

In spite of all the mentioned peculiarities, actual evaluation of
the ensemble average in Eq.(\ref{tot}) can be reduced to the
standard random matrix calculation of the GOE density and as such
can be performed in a closed form for any $N$ in terms of Hermitian
polynomials \cite{YFglass}. This is important, as the same method
actually proved to be useful in other similar problems, as e.g. the
problem of determining the distribution of the maximum of a Gaussian
random field, see \cite{AW}.

Large $N$ asymptotics of the mean number of stationary points can
be extracted from known behavior of Hermite polynomials. The
explicit calculation revealed\cite{YFglass} that the replica
symmetry breaking (interpreted in the standard way as ergodicity
breaking) in the zero-temperature limit of the $R=\infty$ version
of the model with parabolic confinement is accompanied by the
emergence of an exponentially large mean total number of
stationary points in the energy landscape. As common wisdom was to
expect that the low-temperature thermodynamics should be dominated
by minima rather than by the totality of stationary points the
issue called for further investigation, in particular in a general
$R<\infty$ case. In two recent short communications
\cite{BrayDean,FSW} the calculations of \cite{YFglass} were
independently and essentially generalized in a way providing
access to the densities of minima \cite{FSW}, as well as to
stationary points with an arbitrary index \cite{BrayDean}, and at
a given value of the potential, hence energy.

The results in \cite{FSW} were given as a function of both
parameters $\mu$ and $R$. That analysis revealed that generally
the domain of existence of the glassy phase with broken ergodicity
(at zero temperature $T$) turns out to be always associated with
the existence of exponentially many stationary points in the
energy landscape, but not necessarily exponentially many minima.
In an attempt to extend those considerations to finite
temperatures the authors also constructed a simple variational
functional providing an upper bound on the true free energy of the
$\mu>0,R=\infty$ version of the problem. Surprisingly, counting
stationary points in that simple-minded approximation was already
enough to capture such a nontrivial feature as the precise
position of the de-Almeida-Thouless line in the whole $(\mu,T)$
plane.

The aim of the present paper is to describe the random-matrix
evaluation of landscape complexities in a rather general setting.
The calculations are performed for a broad class of Gaussian
landscapes V({\bf x}) with monotonically increasing, concave
 confining potentials, i.e. for the energy functions
\begin{equation}\label{Hgen}
{\cal H}({\bf x})=N\,U\left(\frac{{\bf x}^2}{2N}\right)+V({\bf
x})\,,
\end{equation}
assuming $U'(z)>0,U(z)''\ge 0, \,\,\forall z\ge 0$, with the first
(confining) term chosen in the rotationally-invariant scaling form
ensuring a well-defined large $N$ limit of the model. Solving this
problem will also be used as an opportunity to provide a rather
detailed exposition of the techniques sketched in our earlier
short communication \cite{FSW}. We will be able to treat in
considerable generality the smooth confinement case, and compare
it with non-analytic (hard-wall) confinement studied in \cite{FSW}
(and independently by A Bray and D Dean in \cite{BrayDean}).

Our main results can be briefly summarized as follows. Define the
function
\begin{equation}\label{not}
\mu(z)=U'\left(\frac{z}{2}\right)\,.
\end{equation}
 We find that for a generic, smooth
concave confining potentials $U(z)$ with a continuous
 derivative $\mu(z)$ the condition ensuring that
the mean {\it total} number of stationary points in the energy
landscape is exponentially large in $N$ coincides with one
ensuring that the same is true for the mean number of minima.
Different behaviour of two complexities for a particle confined in
a finite box reported in \cite{BrayDean,FSW} is shown to be a
consequence of non-analyticity of the hard-wall confinement
potential. For a smooth confinement the common condition of
positivity of the corresponding landscape complexities (defined as
the logarithms of the mean total number per degree of freedom $N$)
reads
\begin{equation}\label{zero11}
\mu\left(-\frac{f'(0)}{f''(0)}\right)< \sqrt{f''(0)}\,.
\end{equation}
As shown in Appendix D, such an inequality is precisely the
condition of instability of the zero-temperature replica symmetric
solution of the associated problem in statistical
mechanics\footnote{This result, in particular, gives a kind of
{\it aposteriori} justification of the use of the mean value of
stationary points/extrema as sensible characteristics of
high-dimensional landscapes. Indeed, in the ideal world we would
like to know the typical rather than the mean values, i.e. to
replace "annealed" averages featuring in (\ref{KR}) and
(\ref{KRM}) with the corresponding "quenched" ones. The latter
would require performing the averaging of the {\it logarithm} of
the number of stationary points. Unfortunately, the present level
of art in the field makes such a quenched calculation hardly
feasible.}, see Eq.(\ref{zero}). The corresponding annealed
complexity of minima vanishes cubically when approaching the
critical confinement (\ref{zero1}), whereas the cumulative
annealed complexity vanishes quadratically with the distance to
criticality. Finally, in the appendix E we further investigate,
using the method due to Bray and Dean\cite{BrayDean} the annealed
complexities of stationary points at a fixed value of the {\it
index} ${\cal I}$ (the number of the negative eigenvalues of the
Hessian). The calculation is performed for the simplest case of
parabolic confinement Eq.(\ref{Hpar}), when $\mu(z)=const=\mu$,
hence Eq.(\ref{zero1}) amounts to $\mu<\mu_{cr}=\sqrt{f''(0)}$.
The results reveal that the only stationary points with
non-vanishing annealed complexity arising precisely at the
critical value $\mu=\mu_{cr}$ are those for which the number of
negative directions is not extensive: $\lim_{N\to
\infty}\frac{\cal I}{N}=0$. As we move inside the glassy phase
away from the critical value $\mu=\mu_{cr}$, stationary points
with an increasing range of indices start to have a positive
complexity.

In the Appendices A,B, and C we provide a few useful technical
details, some of them are not immediate to find in the available
literature. We comment on the geometric features of the landscapes
behind the curious formula (\ref{ident}) in the Appendix A, and
discuss the distribution of the diagonal element of the resolvent
of a random GOE matrix in the Appendix B. In the Appendix C we
provide a short overview of a powerful heuristic random-matrix
technique due to Dean and Majumdar \cite{DeanMaj} based on the
functional integration. We found this method indispensable when
calculating the complexity of minima. Finally, a
statistical-mechanics calculation used for extracting the point of
zero-temperature replica symmetry breaking for the present class
of models within the framework of the replica method is sketched
in the Appendix D, and the calculation of complexity of stationary
points with a given (extensive) index is sketched in Appendix E.

{\bf Acknowledgements}. This research was supported by EPSRC grant
EP/C515056/1 "Random Matrices and Polynomials: a tool to
understand complexity". The authors are grateful to H.-J. Sommers
for collaboration at early stages of the project (see papers
\cite{FS}, \cite{FSW}), and to A. Bray and D. Dean for their
stimulating interest in this type of problems. The authors also
appreciate detailed comments of anonimous referees on the first
version of the manuscript which helped much towards improving the
style of the presentation.

\section{Complexity of stationary points versus complexity of
minima}
\subsection{The density of stationary points for
spherically-symmetric potentials} Our goal is to study the density
of stationary points of the function (\ref{Hgen}) around position
${\bf x}$, subject to the condition that the random part $V({\bf
x})$ takes the prescribed value $V$. This is given by a variant of
the Kac-Rice formula (\ref{KR}):
\begin{equation}\label{den1}
 \left\langle\rho_s(V,\x,[U])\right\rangle = \left\langle \left| \det\left[
\frac{\partial^{2}\mathcal{H}}{\partial {\bf x} \partial {\bf
x}}\right]
 \right| \delta\left(\frac{\partial\mathcal{H}({\bf x})}{\partial {\bf x}}\right) \delta[V - V({\bf
x})]\right\rangle\,,
\end{equation}
where we used self-explanatory short-hand notations.

For the choice of the confining potential in Eq.(\ref{Hgen}) we
obviously have for the gradient vector and for the Hessian matrix
\begin{eqnarray}
 \frac{\partial \mathcal{H}}{\partial \x}& = &\x \,\mu\left(\frac{\x^{2}}{N}\right) + \frac{\partial V}{\partial \x},\\
 \frac{\partial^2 \mathcal{H}}{\partial \x \partial \x}& = & \mu\left( \frac{\x^{2}}{N} \right)\,\hat{I}_{N} + \hat{M}(\x) + \frac{\partial^{2} V}{\partial \x \partial \x},
\end{eqnarray}
where we used the function $\mu(z)$ defined according to
(\ref{not}), as well as $N\times N$ rank-one matrix $\hat{M}(\x)$
with entries expressed via components $x_i,\,i=1,\ldots,N$ of
${\bf x}$ as
\begin{equation}\label{rank1}
 [\hat{M}(\x)]_{ij} = \frac{x_{i}x_{j}}{N}\,U''\left( \frac{\x^{2}}{2N}\right)\equiv 2\frac{x_{i}x_{j}}{N}
\mu'\left( \frac{\x^{2}}{N}\right).
\end{equation}
Note that for the parabolic confinement case Eq.(\ref{Hpar})
obviously $\mu(z)\equiv \mu$. We assume in the subsequent analysis
that $\mu(z)$ is non-decreasing: $\mu'(z)\ge 0\,,\forall z\ge 0$.

 We find it also convenient to collect all disorder-dependent factors
in the function
\begin{equation}\label{start}
 \mathcal{F}_s(V,\x,\hat{K},[U]) = \left\langle \delta[V-V(\x)]\,\delta\left[ \x
\,\mu\left(\frac{\x^{2}}{N}\right) + \frac{\partial V}{\partial
\x}\right] \delta\left[\hat{K} - \frac{\partial^{2} V}{\partial \x
\partial \x}\right]\right\rangle
\end{equation}
and use the above expressions to rewrite the density
Eq.(\ref{den1}) in the form
\begin{equation}\label{dena}
 \left\langle\rho_s(V,\x,[U])\right\rangle = \int d\hat{K} \left| \det \left[ \hat{K} + \hat{I}_{N}
\,\mu\left( \frac{\x^{2}}{N} \right) + \hat{M}(\x) \right]
\right|\mathcal{F}(V,\x,\hat{K},[U])\,.
\end{equation}

In the rest of the calculations we assume that the covariance
function $f(x)$ in Eq.(\ref{cov1}) has finite values of the first
two derivatives at the origin: $|f'(0)|<\infty$ and
$|f''(0)|<\infty$.  Actually, from the point of view of
statistical mechanics such a condition ensures the existence of a
nontrivial phase transition at zero temperature, see
Eq.(\ref{zero11}). Note that in the case of long-ranged potentials
behaving like $f(x)\sim x^{\gamma}$ with $0<\gamma<1$ at large $x$
this condition requires imposing some regularization at small
arguments, e.g. $f(x)=(a+x)^{\gamma},\, a>0$.

Let us introduce the following notations for the subsequent use,
remembering $f'(0)<0$:
\begin{equation}\label{9}
\mu_{cr}=\sqrt{f''(0)}\,,\quad R_{cr}=\sqrt{|f'(0)|/f''(0)}\,,
\quad g^2=f''(0)-\frac{f'(0)^2}{f(0)}\ge 0\,.
\end{equation}

  Our analysis of
Eq.(\ref{start}) starts with introducing the Fourier integral
representation for each of the delta-functional measures, - of
scalar, vector, and matrix argument, correspondingly. This step
facilitates performing the ensemble average explicitly as
averaging exponentials containing terms linear with respect to
Gaussian variables requires only the knowledge of their
covariances. In the course of this procedure one has to exploit a
few identities which follow from Eq.(\ref{cov1}) after simple
calculations, as well as from the expressions Eq.(\ref{JPD1}).
Namely,
\begin{equation}\label{4}
\left\langle V^2({\bf
x})\right\rangle=Nf(0),\,\left\langle\frac{\partial V} {\partial
{\bf x}}\frac{\partial V} {\partial {\bf x}}\right\rangle=-f'(0),
\quad \left\langle V\,\frac{\partial V}{\partial {\bf x}}
\right\rangle=\left\langle \frac{\partial V}{\partial {\bf x}}\,
\frac{\partial^2 V}{\partial {\bf x}\partial{\bf
x}}\right\rangle=0,
\end{equation}
\begin{equation}\label{5}
\left\langle\left[\mbox{Tr}\left(\hat{A} \frac{\partial^2
V}{\partial {\bf x}\partial {\bf
x}}\right)\right]^2\right\rangle=\frac{1}{N}f''(0)\left[2\mbox{Tr}\hat{A}^2+
\left(\mbox{Tr}\hat{A}\right)^2\right],
\end{equation}
\begin{equation}\label{6}
\left\langle V({\bf x})\mbox{Tr} \left(\hat{A}\, \frac{\partial^2
V}{\partial {\bf x}\partial{\bf
x}}\right)\right\rangle=f'(0)\mbox{Tr}\hat{A},\quad \left\langle
\left({\bf a} \frac{\partial V}{\partial {\bf
x}}\right)^2\right\rangle=-f'(0)\,{\bf a}^2\,,
\end{equation}
where ${\bf a}$ and $\hat{A}$ are an arbitrary $N-$component
vector and an $N\times N$ real symmetric matrix, respectively.

After these steps one can easily integrate out the scalar and
vector Fourier variables as they appear as simple quadratic terms
in the exponential (the integration over the matrix Fourier
variable requires a little bit more care, and is done at the next
step). Thus, we arrive at the following expression:
\begin{equation}\label{7}
{\cal F}(V,{\bf x},\hat{K},[U])\propto \exp\left\{\frac{{\bf
x}^2}{2f'(0)}\left[\mu\left(\frac{\x^{2}}{N}\right)\right]^2-\frac{V^2}{2Nf(0)}\right\}\,{\cal
I}(\hat{K})
\end{equation}
where ${\cal I}(\hat{K})$ stands for the integral over the
remaining Fourier variables, that is over a real symmetric
$N\times N$ matrix $\hat{P}$, and is given by
\begin{equation}\label{8}
\quad\quad {\cal I}(\hat{K})=
\int\exp\left\{-\frac{\mu_{cr}^2}{N}\mbox{Tr}\hat{P}^2+i\mbox{Tr}\hat{P}
\left(\hat{K}-\frac{f'(0)}{Nf(0)}\,V\hat{I}_N\right)-\frac{g^2}{2N}\left(\mbox{Tr}\hat{P}\right)^2
\right\}\,d\hat{P}\,.
\end{equation}
Here, and henceforth, we will systematically disregard various
multiplicative constant factors for the sake of brevity. They
can always be restored from the normalisation conditions whenever
necessary.

To deal with the remaining integral over the matrix $\hat{P}$ in
(\ref{8}) in the most economic way it is convenient to employ
first a Gaussian integral over an auxiliary scalar variable $t$,
and in this way to "linearize" the last term in the exponential in
Eq.(\ref{8}):
\begin{equation}\label{HS}
e^{-\frac{g^2}{2N}\left(\mbox{\small
Tr}\hat{P}\right)^2}=\left(\frac{N}{2\pi}\right)^{1/2} \int
e^{-\frac{N}{2}t^2\pm i gt\,\mbox{\small Tr}\hat{P}}dt
\end{equation}
Substituting this identity back into (\ref{8}) converts the matrix
integral into a standard Gaussian one which can be immediately
performed, yielding
\begin{eqnarray}\label{8a}
&&{\cal F}(V,{\bf x},\hat{K},[U])\propto \exp\left\{\frac{{\bf
x}^2}{2f'(0)}\left[\mu\left(\frac{\x^{2}}{N}\right)\right]^2-\frac{V^2}{2Nf(0)}\right\}
\\ \nonumber &\times&
\int_{-\infty}^{\infty}
e^{-\frac{N}{2}t^2-\frac{N}{4\mu_{cr}^2}\mbox{Tr}
\left[\hat{K}-\left(gt+\frac{f'(0)}{f(0)}\,\frac{V}{N}\right)\hat{I}_N\right]^2}
\,dt
\end{eqnarray}
Before substituting this expression back into Eq.(\ref{dena}), we
find it convenient to rescale coordinates and the value of the
potential as $ \x \to \sqrt{N} \x$ and $V \to NV$, and to
introduce two matrices $\hat{H}$ and $\mathcal{\hat{M}}$ defined
as
\begin{equation}
\hat{H}=\hat{K}-\left(gt+\frac{f'(0)}{f(0)}\,V\right)\hat{I}_N
\end{equation}
and
\begin{equation}\label{rank1b}
\mathcal{\hat{M}} = s \hat{I}_{N} +2\,
\frac{\mu'(\x^2)}{\mu_{cr}}\,\x\otimes \x^T\,,
\end{equation}
where we introduced a short-hand notation
\begin{equation}\label{seff}
s=\frac{1}{\mu_{cr}}\left[\mu\left(\x^{2}\right) + gt +
\frac{f'(0)}{f(0)}V\right]\,.
\end{equation}
and the diadic product $\x\otimes \x^T$ stands for the (rank-one)
$N\times N$ matrix with components $x_ix_j$.

 As a result, we find that the mean density of stationary
points Eq.(\ref{dena}) is expressed in terms the rescaled
variables as
\begin{eqnarray}\label{denb}
&&\rho(V,\x,[U]) = \mathcal{N}_{s}\exp N\left(
\frac{\x^{2}}{2f'(0)}\mu^2\left(\x^{2}\right) -
\frac{V^{2}}{2f(0)}\right)\int dt e^{-Nt^{2}/2}{\cal D}_s(t
,V,[U])
\end{eqnarray}
\begin{eqnarray}
 &&{\cal D}_s(t,V,[U])= \left\langle \left|
\det(\hat{H}+\mathcal{\hat{M}})
 \right|\right\rangle_{GOE}\label{denfac}
\end{eqnarray}
where ${\cal N}_s$ is the required normalisation constant and
$\langle \ldots \rangle_{GOE}$ denotes the average with respect to
the standard GOE measure $\mathcal{P}(\hat{H})\,d\hat{H}\propto
\exp(-N\mbox{ Tr}\hat{H}^{2}/4)\,d\hat{H}$. Now we can use the
rank-one character of the second term in (\ref{rank1b}) which
implies the following identity
\begin{equation}\label{rank1a}
\det(\hat{H}+\mathcal{\hat{M}})=\left[1+2\,
\frac{\mu'(\x^2)}{\mu_{cr}}\,G_H(\x)\right]\det\left(\hat{H}+s\hat{I}\right)
\end{equation}
where
\begin{equation}\label{reso}
G_H(\x)={\bf x}^T\,\frac{1}{\hat{H}+s\,\hat{I}}\,\x
\end{equation}
is a diagonal element of the resolvent of the random matrix
$\hat{H}+s\,\hat{I}$. Note that with the parameter $s$ being real
as required by the identity (\ref{rank1a}), the resolvent can take
arbitrary large values when an eigenvalue of the random matrix $H$
occurs close enough to $-s$. As in the large $N$ limit the
eigenvalues of GOE matrices fill in densely the interval
$(-2,2)$\cite{Mehta} we can expect the resolvent to retain strong
fluctuations for $|s|<2$. Outside that interval however we may
expect self-averaging of the resolvent. The explicit calculation
performed in the Appendix B confirms this intuition, and
establishes the magnitude of the fluctuations inside the spectrum.
Namely, we show that for $N\to \infty$ the distribution function
of $G_H(\x)$ tends to the Cauchy law centered around the mean
value $\langle G_H(\x)\rangle_{GOE}=\frac{1}{2}s\,\x^2$ and with
the widths $\Gamma=\left\langle|G_H(\x)-\langle
G_H(\x)\rangle_{GOE}|\right\rangle_{GOE}=\frac{1}{2}\sqrt{4-s^2}$
for $|s|<2$, and $\Gamma=0$ otherwise. We therefore conclude that
the first factor in Eq.(\ref{rank1a}) is typically of the order of
unity at every realisation of the disorder. Recalling that only
factors of order $\exp\{O(N)\}$ in the mean density of stationary
points are relevant for calculating the complexity, we can safely
disregard the mentioned factor in the rest of the calculation by
replacing Eq.(\ref{denfac}) with
\begin{equation}\label{denfac1}
{\cal D}_s(t,V,[U])= \left\langle \left| \det(\hat{H}+s\,\hat{I})
 \right|\right\rangle_{GOE}
\end{equation}
and remembering the relation (\ref{seff}) of $s$ to parameters of
the problem.

 For the asymptotic analysis of complexity we
further use the result proved in \cite{YFglass}. It essentially
claims that the expectation of the determinant in the right-hand
side of Eq.(\ref{denfac1}) can be represented in the large $N$
limit as
\begin{equation}\label{11}
{\cal D}_s(t,V,[U])\propto \exp{N\Phi\left(s\right)}
\end{equation}
where the function $\Phi(s)=\Phi(-s)$ is given explicitly by
\begin{equation}\label{12a}
\Phi(s\ge
0)=\frac{s^2}{4}-\theta(s-2)\left[\frac{s\,\sqrt{s^2-4}}{4}+\ln{\left(\frac{s-\sqrt{s^2-4}}{2}\right)}\right]
\end{equation}
The method used in \cite{YFglass} was based on relating  ${\cal
D}_s(t,V,[U])$ to the mean eigenvalue density of GOE, and
subsequent lengthy saddle-point analysis of an integral
representation for that density. Perhaps, a shorter way to
understand the above asymptotics is to notice that actually,
\begin{equation}\label{12b}
\Phi(s)=\,\int_{-2}^2\ln{|s+\lambda|}\,\rho_{sc}(\lambda)\,d\lambda
,\quad \rho_{sc}(\lambda)=\frac{1}{2\pi}\sqrt{4-\lambda^2}
\end{equation}
where the integral is understood in the sense of the principal
value. The above formula can be verified by applying ideas from
statistical mechanics to the evaluation of $D_s(t,V)$. A
mathematically rigorous method of analysing this type of problems
can be found in a detailed paper by Boutet de Monvel, Pastur and
Shcherbina \cite{BPS}. An alternative, and very transparent way of
heuristic asymptotic analysis based on the concept of a functional
integral was proposed in a recent insightful work by Dean and
Majumdar \cite{DeanMaj}. We provide an overview of the
Dean-Majumdar approach in the Appendix C.

We thus arrive to the following expression for the density of
stationary points
\begin{equation}\label{denstat}
 \rho(V,\x,[U]) \sim \mathcal{N}_{s}\exp\left(N\left[\frac{\x^{2}}{2f'(0)}\mu^2(\x^{2}) -
 \frac{V^{2}}{2f(0)}\right] \right)\int_{-\infty}^{\infty} dt e^{N[\Phi(s) - t^{2}/2]}
\end{equation}
valid in the large-$N$ limit up to factors of order of unity.

\subsection{The cumulative complexity of stationary points for a smooth confinement}

The mean total number of the stationary points for a given shape
of the spherically-symmetric confining potential is obtained by
further integrating this density over the position $\bf x$ and
over the values of the potential $V$:
\begin{eqnarray}
&&\langle\#_s\rangle=\int_{\mathbb{R}^N}d\x\int_{-\infty}^{\infty}\,\rho(V,\x,[U])\,
dV
\end{eqnarray}
It is natural to use the fact that the integrand is spherically
symmetric by passing to the integration over the radial variable
$q=\x^2$ and the angular coordinates $\Omega$, so that $d\x\propto
q^{(N-2)/2}dq d\Omega$. As a result, we have
\begin{eqnarray}\label{snumber}
&&\langle\#_s\rangle\propto \int_{0}^{\infty}\,e^{N{\cal
L}(q)}\,\frac{dq}{q} {\cal F}(q),\quad {\cal F}(q)=
\int_{-\infty}^{\infty} \int_{-\infty}^{\infty} e^{\,-N{\cal
S}[V,q,t]}\,dt\, dV
\end{eqnarray}
where
\begin{equation}\label{action}
 {\cal
S}(V,q,t)=\frac{V^2}{2f(0)}+\frac{t^{2}}{2}-\Phi(s),\quad {\cal
L}(q)=\frac{q}{2f'(0)}\mu^2(q)+\frac{1}{2}\ln{q}\,,
\end{equation}
with (cf. (\ref{seff}))
\begin{equation}\label{seff1}
s=\frac{1}{\mu_{cr}}\left[\mu\left(q\right) + gt +
\frac{f'(0)}{f(0)}V\right]
\end{equation}
and $\Phi(s)$ given by Eq.(\ref{12a}). As usual, we disregarded in
(\ref{snumber}) all constant proportionality factors which will be
restored at a later stage from the normalisation condition.

 For our analysis of this expression in the limit $N\gg 1$
we find it convenient to evaluate first the integrals over $t,V$
with the Laplace method.  We find ${\cal F}(q) \propto
\exp{[-N{\cal S}(V_*,t_*)]}$, with $V_*$ and $t_*$ being the
values minimizing ${\cal S}(V,q,t)$ for a given $q$, and
satisfying the system of equations:
\begin{equation}\label{14}
V_*=\frac{f'(0)}{\mu_{cr}}\frac{d\Phi(s)}{ds}|_{s=s_*},\quad
t_*=\frac{g}{\mu_{cr}}\frac{d\Phi(s)}{ds}|_{s=s_*},
\end{equation}
 where according to Eq.(\ref{seff1})
\begin{equation}\label{15}
s_*=\frac{1}{\mu_{cr}}[\mu(q)+gt_*+\frac{f'(0)}{f(0)}\,V_*]\equiv
\frac{1}{\mu_{cr}}[\mu(q)+\frac{\mu_{cr}^2}{f'(0)}\,V_*]\,,
\end{equation}
the second equality following from the obvious relation
$t_*=gV_*/f'(0)$ implied by Eq.(\ref{14}), and the definition of
$g^2$ in Eq.(\ref{9}). Moreover, for $s_*\ge 0$ Eq.(\ref{12a})
implies that
\begin{equation}\label{16}
\frac{d\Phi(s)}{ds}|_{s=s_*}=\frac{1}{2}\left[s_*-\theta(s_*-2)\sqrt{s_*^2-4}\right]
\end{equation}
Assuming first $0\le s_*\le 2$, the first of the equations
(\ref{14}) together with Eqs.(\ref{15},\ref{16}) immediately
yields
\begin{equation}\label{less2}
V_*=\mu(q) \frac{f'(0)}{\mu_{cr}^2}, \quad
s_*=2\mu(q)\frac{1}{\mu_{cr}}, \quad\mbox{and}\,\,t_*=\mu(q)
\frac{g}{\mu_{cr}^2}
\end{equation}
Obviously, this solution is compatible with our assumption $0\le
s_*\le 2$ only as long as $0\le \mu(q)\le\mu_{cr}$. On the other
hand, assuming $s_*>2$ the first of the equations (\ref{14})
solved together with Eq.(\ref{16}) yields the relation
\begin{equation}\label{ss}
s_*=\frac{\mu_{cr}}{f'(0)}V_*+\left(\frac{\mu_{cr}}{f'(0)}V_*\right)^{-1}\,
,\end{equation} which is consistent with our assumption.
Substituting for $s_{*}$ from Eq.(\ref{15}) one immediately finds
\begin{equation}\label{more2}
V_*=\frac{f'(0)}{\mu(q)},\quad t_*=\frac{g}{\mu(q)}\quad
\mbox{for}\,\, \mu(q)>\mu_{cr}\,.
\end{equation}
In fact, $V_*$ is nothing else but the most probable value of the
potential $V({\bf x})$ at a stationary point of the energy
surface ${\cal H}({\bf x})$ situated at the distance
$|\x|=\sqrt{NR}$ from the origin. We see that $V_*$ experiences a
a drastic change in its behaviour at the values of radial
parameter $q$ at which the curve $\mu(q)$ crosses the value
$\mu_{cr}=\sqrt{f''(0)}$ and zero. To simplify our analysis we
assume in the following that $\mu(q)>0$ and $\mu'(q)\ge 0,
\,\forall q\ge 0$, i.e. that $\mu(q)$ is a non-decreasing function
of its argument. Then the equation $\mu(q)=\mu_{cr}$ has either a
single solution $q=q_*$ when $\mu(0)<\mu_{cr}<\mu(q\to\infty)$, or
no solution at all (this happens if $\mu_{cr}<\mu(0)$, or
$\mu_{cr}>\mu(\infty)$.

The corresponding values of ${\cal S}(V_*,t_*,q)$ can be found by
substituting (\ref{less2}) and (\ref{more2}) into
Eq.(\ref{action}). In particular, the expression (\ref{ss}) in the
regime (\ref{more2}) implies the identity
$\frac{1}{2}[s_*-\sqrt{s_*^2-2}]=\frac{\mu_{cr}}{\mu(q)}$, so that
$\Phi(s_*)=\frac{\mu_{cr}^2}{2\mu^2}+\frac{1}{2}-\ln{\frac{\mu_{cr}}{\mu(q)}}$,
 and we obtain after some simple algebra:
\begin{equation}\label{18}
{\cal S}(V_*,t_*,q)=\left\{\begin{array}{l}
-\frac{1}{2}+\ln{\left(\frac{\mu_{cr}}{\mu(q)}\right)},\quad \mu(q)\ge \mu_{cr}\\
-\frac{\mu^2(q)}{2\mu_{cr}^2}\quad, \,\quad\quad\quad \mu(q)<
\mu_{cr}\end{array}\right.
\end{equation}

Substituting ${\cal F}(q) \propto \exp{[-N{\cal S}(V_*,t_*,q)]}$
back to the integral (\ref{snumber}), we see that the total number
of stationary points is given by
\begin{equation}\label{stp1}
 \langle \#_{s} \rangle= \mathcal{N}_{s}\int_{0}^{\infty}\,e^{\frac{N}{2} \Psi(q)}\,\frac{dq}{q}\,,
\end{equation}
where ${\cal N}_s$ encapsulates all the necessary normalisation
factors, and
\begin{equation}\label{psi1}
 \Psi(q) = \begin{cases} \frac{q\mu^2(q)}{f'(0)}+1 -
\ln\left( \frac{\mu^2_{cr}}{\mu^2(q)}\right) + \ln q, &\text{for $\mu(q) \geq \mu_{cr}$,}\\
            \frac{q\mu^2(q)}{f'(0)} +\frac{\mu^2(q)}{\mu_{cr}^{2}} +
\ln q, &\text{for $\mu(q)<\mu_{cr}$.}
           \end{cases}
\end{equation}
To find $\mathcal{N}_{s}$ it suffices to consider the special
case where $\mu(q)\to\infty$ when only the minimum at the origin
survives and hence $\langle \#_{s}\rangle \to 1 $. This condition
fixes with exponential in $N$ accuracy
\begin{equation}\label{norma}
\mathcal{N}_{s} \sim e^{-N\ln R_{cr}}
\end{equation}
where the scale $R_{cr}$ was defined in Eq.(\ref{9}). Taking this
factor into account, and assuming that the equation
$\mu(q)=\mu_{cr}$ has a single solution $q=q_*$, we split the
integration range in (\ref{stp1}) into two natural pieces $0<q\le
q_{*}$ and $q>q_*$. Then the exponential in $N$ part of the mean
number of stationary points can be found from
\begin{equation}\label{stp2}
 \langle \#_{s} \rangle=\int_{0}^{q_*}\,e^{\frac{N}{2}
\Psi_{<}(q)}\,\frac{dq}{q} +\int_{q_*}^{\infty}\,e^{\frac{N}{2}
\Psi_{>}(q)}\,\frac{dq}{q}\,,
\end{equation}
where remembering $f'(0)=-\mu^2_{cr}R_{cr}^2$ we defined
\begin{equation}\label{psi2}
 \begin{cases}\Psi_{>}(q)=1-w + \ln{w}, \,\, \text{where}\,\,
w=\frac{q\mu^2(q)}{\mu^2_{cr}R^2_{cr}} , &\text{for $\mu(q) \geq \mu_{cr}\equiv \mu(q_*)$,}\\
         \Psi_{<}(q)=w\frac{R^2_{cr}}{q}\left(1-\frac{q}{R_{cr}^2}\right)+
\ln \left(\frac{q}{R_{cr}^2}\right), &\text{for
$\mu(q)<\mu_{cr}\equiv \mu(q_*)$.}
           \end{cases}
\end{equation}
As the maximum of the function $\Psi_{>}(q)=1-w + \ln{w}$ is
achieved at $w=1$ and equal to zero we have $\Psi_{>}(q)\le
0,\,\forall q>q_{*}$, and hence only the first integral in
(\ref{stp2}) can yield a positive complexity
\begin{equation}\label{compls}
\Sigma_{s}=\lim_{N\to
\infty}\frac{1}{N}\ln{\langle\#_s\rangle}=\lim_{N\to
\infty}\frac{1}{N}\ln{\left(\int_{0}^{q_*}\,e^{\frac{N}{2}
\Psi_{<}(q)}\,\frac{dq}{q}\right)}
\end{equation}
To extract the large-$N$ asymptotics of the latter integral we
calculate
\begin{equation}\label{deriv}
\frac{d\Psi_{<}(q)}{dq}=\frac{dw}{dq}\left(\frac{R_{cr}^2}{q}-1\right)+\frac{1}{q}
\left(1-\frac{\mu^2(q)}{\mu_{cr}^2}\right),\quad \mu(q)<
\mu(q_*)=\mu_{cr}\,.
\end{equation}

The analysis below proceeds separately for two cases:
$0<q_*<R_{cr}^2$ and $0<R_{cr}^2<q_*$.
\begin{enumerate}
\item Let us first consider $q_*<R_{cr}^2$, with the goal to demonstrate
that for such a choice there is no contribution to positive
complexity. We have $\frac{dw}{dq}>0$, so that the derivative in
Eq.(\ref{deriv}) is obviously positive:
$\frac{d\Psi_{<}(q)}{dq}>0,\,q<q_*$. Using that
$w_*=w(q_*)=\frac{q_*}{R^2_{cr}}<1$ we see
$\Psi_{<}(q)<\Psi_{<}(q_*)=1-w_* + \ln{w_*}<0$ and conclude that
in such a case there is no exponentially large contribution to the
integral, hence no positive complexity.
\item
Assume now $0<R_{cr}^2<q_*$. We shall demonstrate the existence of
a point $q_0\in [R^2_{cr},q_*]$ where the function $\Psi_{<}(q)$
attains a positive maximum $\Psi_{<}(q_0)>0$. This in turn will
imply a positive complexity.

To begin with, $R_{cr}^2<q_*$ implies $\mu(R_{cr}^2)<
\mu(q_*)=\mu_{cr}\,$ as the function $\mu(q)$ is non-decreasing.
 Then (\ref{deriv}) implies
$\frac{d\Psi_{<}(q)}{dq}|_{q=R_{cr}^2}=\frac{1}{R_{cr}^2}
\left(1-\frac{\mu^2(R_{cr}^2)}{\mu_{cr}^2}\right)>0$. Note that
from (\ref{psi2}) follows $\Psi_{<}(q<R_{cr}^2)<0$, and
$\Psi_{<}(q=R_{cr}^2)=0$. Therefore $\Psi_{<}(q)>0$ in some right
vicinity of $q=R^2_{cr}$.

On the other hand, we have already seen above that
$\Psi_{<}(q_*)<0$ and also (\ref{deriv}) implies
\begin{equation}\label{deriv1}
\frac{d\Psi_{<}(q)}{dq}|_{q=q_*}=\frac{dw}{dq}|_{q_*}\left(\frac{q_{cr}^2}{q_*}-1\right)<0\,.
\end{equation}
Then by continuity there must exists a point $q=q_1\in
(R^2_{cr},q_*)$ such that $\Psi_{<}(q_1)=0$. Remembering
$\Psi_{<}(q=R_{cr}^2)=0$ and positivity of $\Psi_{<}(q)$ to the
immediate right of $q=R_{cr}^2$ we conclude on the existence of
another point $q=q_0\in{(R^2_{cr},q_1)}$ where the function
$\Psi_{<}(q)$ attains at least one positive maximum
$\Psi_{<}(q_0)>0$.
\end{enumerate}

Combining all these facts, we see that under these conditions
Eq.(\ref{compls}) leads to the positive cumulative complexity of
stationary points:
\begin{equation}\label{complsfin}
\Sigma_{s}=\frac{1}{2}\max_{q\in(R^2_{cr},q_*)}\left[\frac{\mu^2(q)}{\mu^2_{cr}}\left(1-\frac{q}{R_{cr}^2}\right)+
\ln \left(\frac{q}{R_{cr}^2}\right)\right]>0,
\end{equation}
as long as $\mu(R_{cr}^2)<\mu_{cr}$. Taking into account
definitions (\ref{9}), the latter condition can be written in
terms of the covariance function of the random potential as
\begin{equation}\label{zero1}
\mu\left(-\frac{f'(0)}{f''(0)}\right)< \sqrt{f''(0)}\,.
\end{equation}
This inequality is nothing else but precisely the condition of
instability of the zero-temperature replica symmetric solution of
the associated problem in statistical mechanics, see
Eq.(\ref{zero}) in the Appendix D.

For the simplest case of a parabolic confinement (\ref{Hpar}) with
$\mu(q)=\mu<\mu_{cr}$ we can easily find the cumulative complexity
as an explicit function of $\mu$. Indeed, the point where
$\Psi_<(q)$ attains its maximum is given by
$q_0=R_{cr}^2\frac{\mu^2_{cr}}{\mu^2}$, and
\begin{equation}\label{complsmu}
\Sigma_{s}=\frac{1}{2}\left[-1-\ln{\frac{\mu^2}{\mu_{cr}^2}}+\frac{\mu^2}{\mu_{cr}^2}\right]\ge
0,\quad \mbox{for}\,\,\mu\le \mu_{cr}
\end{equation}
in full agreement with the result reported earlier in
\cite{YFglass}. When approaching the critical value $\mu=\mu_{cr}$
the cumulative complexity vanishes quadratically:
$\Sigma_{s}\approx\left(1-\frac{\mu}{\mu_{cr}}\right)^2$.

Let us show that such type of critical behaviour is generic.
Introducing the short-hand notations  $\tilde{q}=q/R_{cr}^2$ and
$\tilde{\mu}(\tilde{q})=\frac{\mu(q)}{\mu_{cr}}$, we rewrite
$\Psi_{<}(q)$ as
\begin{equation}\label{psis}
\Psi_{<}(\tilde{q})=(1-\tilde{q})\tilde{\mu}^2(\tilde{q})+\ln{\tilde{q}}\,.
\end{equation}
The condition of criticality (\ref{zero1}) is simply
$\mu(R^2_{cr})/\mu_{cr}=1$, which amounts in new notations to
$\tilde{\mu}(1)=1$. Hence we introduce the parameter
$\delta=1-\tilde{\mu}(1)$ which controls the distance to the
criticality. Obviously, the interval $R^2_{cr}\le q\le q_*$
shrinks to zero when $\delta\to 0$, so everywhere in the critical
region $\tilde{q}-1=\epsilon\ll 1$, and we can approximate the
function $\tilde{\mu}(\tilde{q})$ by first two terms in Taylor
expansion: $\tilde{\mu}(\tilde{q})\approx \tilde{\mu}(1)+\epsilon
\tilde{\mu}'(1)\equiv 1-\delta+\epsilon \tilde{\mu}'(1)$.
Substituting this to (\ref{psis}) we obtain
\begin{equation}\label{psisa}
\Psi_{<}\left(\tilde{q}=1+\epsilon\right)=\left(2\delta+O(\delta^2)\right)\epsilon-
\left(\frac{1}{2}+2\tilde{\mu}'(1)+O(\delta)\right)
\epsilon^2+O(\epsilon^3)\,.
\end{equation}
To find the complexity, we should maximize this over $\epsilon$. To
the leading order in $\delta$ the maximum is attained at
$\epsilon_0=\delta/\left(\frac{1}{2}+2\tilde{\mu}'(1)\right)$ and
the corresponding complexity is given by
\begin{equation}\label{psisb}
\Sigma_s=\frac{1}{2}\Psi_{<}\left(\tilde{q}=1+\epsilon_0\right)=
\frac{\delta^2}{1+4\tilde{\mu}'(1)}\,,
\end{equation}
which indeed always vanishes quadratically at criticality.

\subsection{The density of minima for smooth spherically-symmetric
potentials, and the associated complexity}

If one is interested in calculating the density of only minima,
one has to perform the manipulations identical to those in the
first half of the preceding section leading to Eqs.(\ref{denb})
and (\ref{denfac1}), but using Eq.(\ref{KRM}) rather than
(\ref{KR}) as a starting point. Repeating all the steps, one
arrives then to
\begin{eqnarray}\label{minb}
&&\rho_m(V,\x,[U]) = \mathcal{N}_{m}\exp N\left(
\frac{\x^{2}}{2f'(0)}\mu^2\left(\x^{2}\right) -
\frac{V^{2}}{2f(0)}\right)\int dt e^{-Nt^{2}/2}{\cal D}_m(t
,V,[U])
\end{eqnarray}
with
\begin{equation}\label{minfac1}
{\cal D}_m(t,V,[U])\equiv D_m(s)= \left\langle
\det(\hat{H}+s\,\hat{I})\,
\theta\left[\det(\hat{H}+s\,\hat{I})\right]\right\rangle_{GOE}\,,
\end{equation}
where the variable $s$ is given in terms of $t$ and $V$ by the
same expression (\ref{seff}), and ${\cal N}_m$ is a relevant
normalisation constant. As before we use the notation $\langle
\ldots \rangle_{GOE}$ for the average with respect to the standard
GOE measure $\mathcal{P}(\hat{H})\,d\hat{H}\propto \exp(-N\mbox{
Tr}\hat{H}^{2}/4)\,d\hat{H}$. To extract large-$N$ behaviour of
the function $D_m(s)$ we employ again the Dean-Majumdar method as
explained in the Appendix C. In this way one finds that the
function $D_m(s)$ with the required accuracy is given
asymptotically by
\begin{equation}
 D_{m}(s) \propto \begin{cases} \exp \left\{-\frac{N^2}{2}G_{m}(s)+N T_m(s)\right\}  & \text{if $s<2$,} \\
\exp [N \phi(s)]  & \text{if $s>2$,}
            \end{cases}
\end{equation}
where the calculation sketched in the Appendix C gives
\begin{multline}\label{Gmin}
G_{m}(s)=
\frac{1}{216}\left(72s^2-s^4-30s\sqrt{12+s^2}-s^3\sqrt{12+s^2}\right)
-\ln{\frac{(s+\sqrt{s^2+12})}{6}},
\end{multline}
so that $G_{m}(2)=0$, and $\phi(s)$ is the same as expression
(\ref{12a}) for $s>2$, i.e.
\begin{equation}\label{12aa}
\phi(s) = \frac{s^{2}}{4} - \left[\frac{s\sqrt{s^2-4}}{4} +
\ln\left(\frac{s-\sqrt{s^2-4}}{2}\right)\right].
\end{equation}
Explicit expression for the function $T_m(s)$ can be easily found
from the procedure described in the Appendix C, but is immaterial
for our purposes, apart from the fact that continuity of ${\cal
D}(s)$ at $s=2$ and the property $G_m(2)=0$ implies
$T_m(2)=\phi(2)=1$.

 Correspondingly, the density of minima in Eq.(\ref{minb}) is
calculated as the sum of two contributions:
\begin{multline}\label{rho1}
\rho_{m}^{(1)} = {\cal N}_m\exp N\left(
\frac{\x^{2}}{2f'(0)}\mu^2\left(\x^{2}\right)-
\frac{V^{2}}{2f(0)}\right)\\ \times
\int_{-\infty}^{\infty}\exp\left(-\frac{Nt^{2}}{2}+N\,T_m(s)-\frac{N^{2}}{2}G_{m}(s)
 \right)\,\theta(2-s)\,dt\,,
\end{multline}
\begin{multline}\label{rho2}
\rho_{m}^{(2)} = {\cal N}_m \exp N\left(
\frac{\x^{2}}{2f'(0)}\mu^2\left(\x^{2}\right) -
\frac{V^{2}}{2f(0)}\right)\\ \times
\int_{-\infty}^{\infty}\exp\left(-\frac{N
t^2}{2}+N\phi(s)\right)\,\theta(s-2)\,dt\,.
\end{multline}
 To shorten our analysis in
the large-$N$ limit we recall that our final goal is rather to
calculate the exponential in $N$ contribution to the mean number
of minima. As in this process we integrate the above density over
the value of the potential $V$, and over the coordinates $\x$ in
the sample, we can search simultaneously for the optimal values of
variables $t$ and $V$. Let us start with the asymptotic analysis
of the second contribution, Eq.(\ref{rho2}). The shape of the
integrand in this case coincides with that in the analysis of the
density of stationary points for $s>2$, hence the optimal values
are given by (see (\ref{more2}))
\begin{equation}\label{more2a}
V_*=\frac{f'(0)}{\mu(q)},\quad t_*=\frac{g}{\mu(q)}\quad
\mbox{for}\,\, \mu(q)>\mu_{cr}\,.
\end{equation}
where as before we introduced $q=\x^2$. In particular, the most
probable values $V_*$ of the potential $V({\bf x})$ taken over all
the stationary point, or over only minima coincide in this regime.
The corresponding contribution to the density of minima is given
by
\begin{equation}\label{rho2a}
\rho_{m}^{(2)} \propto \exp \frac{N}{2}\left(
\frac{q}{2f'(0)}\mu^2\left(q\right)
-\ln{\frac{\mu_{cr}^2}{\mu^2(q)}}\right).
\end{equation}
This is again the same as contribution of the density of all
stationary points in the regime $\mu(q)>\mu_{cr}$. As we know from
previous analysis , in this regime there can be no exponentially
many stationary points, even less so minima. Similar consideration
also makes it clear that the overall normalization factor ${\cal
N}_m$ (common to both $\rho_{m}^{(1)}$ and $\rho_{m}^{(2)}$) in
the final expression should be (within exponential accuracy) the
same as one used for the total density of stationary points in
Eq.(\ref{norma}).

Now we consider the contribution coming from Eq.(\ref{rho1}),
which can be conveniently rewritten using the relation between $t$
and $s$ as
\begin{multline}\label{rho1a}
\rho_{m}^{(1)} \propto \exp N\left(
\frac{q}{2f'(0)}\mu^2\left(q\right)- \frac{V^{2}}{2f(0)}\right)\\
\times \int_{-\infty}^{2}\exp{-N\left(\frac{1}{2g^2}
\left[\mu_{cr}s-\mu(q)-\frac{f'(0)}{f(0)}\,V\right]^2-T_m(s)+\frac{N}{2}G_{m}(s)\,.
 \right)}\,\,ds\,.
\end{multline}

When $N\to \infty$, the integral over $s$ will be obviously
dominated by the value which minimizes the function $G_m(s)$. One
can satisfy oneself that this minimum occurs precisely at the
boundary of the integration region $s=2$. Defining $s=2-z$ we
expand for small $z$ as $ G_{m}(2-z)\approx \frac{z^{3}}{12} +
\mathcal{O}(z^{4})$, and after substituting this expansion back to
Eq.(\ref{rho1a}) integrate out $z$. This procedure yields an
irrelevant pre-exponential factor, and using $\phi(2)=1$ the
exponential in $N$ contribution to the density takes the form
\begin{equation}\label{rho1b}
\rho_{m}^{(1)} \propto \exp \frac{N}{2}\left(
\frac{q}{f'(0)}\mu^2\left(q\right)-
\frac{V^{2}}{f(0)}+2-\frac{1}{g^2}
\left[2\mu_{cr}-\mu(q)-\frac{f'(0)}{f(0)}\,V\right]^2
 \right),\,
\end{equation}
Next we find that this expression is maximized at the value of the
potential given by
\begin{equation}\label{rho1c}
V_*=\frac{f'(0)}{\mu_{cr}}\,\left[2-\frac{\mu(q)}{\mu_{cr}}\right]\,,
\end{equation}
and the value of the density $\rho_{m}^{(1)}$ at this maximum is
proportional to
\begin{equation}\label{rho1d}
\rho_{m}^{(1)} \propto \exp \frac{N}{2}\left(
\frac{q}{f'(0)}\mu^2\left(q\right)+2-
\left[2-\frac{\mu(q)}{\mu_{cr}}\right]^2
 \right)\,.
\end{equation}

Now this expression can be used for extracting the exponential in
$N$ contribution to the total mean number $\langle \#_{m} \rangle$
of minima, hence the corresponding complexity. This amounts to
multiplying the density (\ref{rho1d}) with the "volume" factor
$q^{(N-2)/2}dq$, integrating over the radial variable $R$ in the
range up to $q=q_*$, such that $\mu(q)<\mu_{cr}$ for $q<q_*$, and
finally multiplying with the overall normalisation factor
$e^{-N\ln R_{cr}}$. Taking the logarithm yields the complexity of
minima
\begin{eqnarray}\label{complm}
\Sigma_{m}&=&\lim_{N\to
\infty}\frac{1}{N}\ln{\langle\#_m\rangle}=\lim_{N\to
\infty}\frac{1}{N}\ln{\left(\int_{0}^{R_*}\,e^{\frac{N}{2}
\Psi_{m}(q)}\,\frac{dq}{q}\right)}\\ &&
=\frac{1}{2}\max_{q\in(0,q_*)}\Psi_{m}(q)
\end{eqnarray}
where \begin{equation}\label{psim} \Psi_m(q)=
2+\ln{\frac{q}{R^2_{cr}}}-\frac{q}{R^2_{cr}}
\frac{\mu^2(q)}{\mu^2_{cr}}-\left[2-\frac{\mu(q)}{\mu_{cr}}\right]^2\,.
\end{equation}

Let us first consider the simplest case of a parabolic confinement
$\mu(q)=\mu<\mu_{cr}$, see (\ref{Hpar}). The  complexity of minima
can be easily found as the function of $\mu/\mu_{cr}$. Again, the
point $q_0$ at which $\Psi_m(q)$ attains its maximum is equal to
the same value $q_0=R_{cr}^2\frac{\mu^2_{cr}}{\mu^2}$ which
delivered earlier the maximum to the function $\Psi_<(R)$, and
\begin{equation}\label{complmmu}
\Sigma_{m}\left(\frac{\mu}{\mu_{cr}}\right)=\frac{1}{2}\left[-3-\ln{\frac{\mu^2}{\mu_{cr}^2}}
+4\frac{\mu}{\mu_{cr}}-\frac{\mu^2}{\mu_{cr}^2}\right],\quad
\mbox{for}\,\, \mu\le \mu_{cr}
\end{equation}
It is evident that
$\Sigma_{m}\left(\frac{\mu}{\mu_{cr}}=1\right)=0$, and it is easy
to check that $d\Sigma_m/d\mu<0$, hence
$\Sigma_{m}\left(\frac{\mu}{\mu_{cr}}=1\right)>0$ for any
$\frac{\mu}{\mu_{cr}}<1$. Close to the critical value
$\frac{\mu}{\mu_{cr}}=1$ the complexity $\Sigma_{m}$ vanishes
cubically: $\Sigma_{m}\approx
\frac{1}{3}\left(1-\frac{\mu}{\mu_{cr}}\right)^3$. This is a
faster decrease in comparison with the quadratic behaviour
demonstrated by the cumulative complexity from (\ref{complsmu}).
In fact note that for any $\mu\le \mu_{cr}$ the inequality
$\Sigma_{s}-\Sigma_{m}=\left(1-\frac{\mu}{\mu_{cr}}\right)^2>0$ holds.

In a general case $\mu'(q)>0$ the point $q_0$ at which the
function $\Psi_<(q)$ from (\ref{psi2}) attains its maximum does
not necessarily coincide with one which delivers the maximum to
$\Psi_m(q)$ from (\ref{psim}). Indeed, the two functions satisfy
the following relation:
\begin{equation}\label{rel}
\Psi_m(q)=\Psi_<(q)-2\left[1-\frac{\mu(q)}{\mu_{cr}}\right]^2\,,
\end{equation}
which implies for derivatives
\begin{equation}\label{rela}
\frac{d}{dq}\Psi_m(q)=\frac{d}{dq}\Psi_<(q)+4\frac{\mu'(q)}{\mu_{cr}}\left[1-\frac{\mu(q)}{\mu_{cr}}\right]\,,
\end{equation}
Hence in the regime $\mu(q)<\mu_{cr}$ interesting for us here
$\frac{d}{dq}\Psi_m(q)>0$ at the point of maximum of the function
$\Psi_<(q)$.

As discussed in the previous section, for $q_\star<R_{cr}^2$ the
derivative $\frac{d}{dq}\Psi_<(q)$ is strictly positive, and thus
the derivative $\frac{d}{dq}\Psi_m(q)$ cannot be zero. This
implies that $\Psi_m(q)$ can only have a maximum if the inequality
$q_*>R^2_{cr}$ holds, and the point of maximum must belong to the
interval $q\in[R^2_{cr},q_*]$. Below we shall prove that in fact
the condition $q_*>R^2_{cr}$ is enough to ensure that complexity
of minima is positive. Let us however note that the simple
arguments used to prove the positivity of the maximum of
$\Psi_s(R)$, hence the positivity of the cumulative complexity
$\Sigma_s$, are not immediately applicable in the case of the
complexity of minima. Indeed, although we have
$\frac{d}{dq}\Psi_m(q)|_{R=R_{cr}^2}=\left[1-\left(\frac{\mu(R^2_{cr})}{\mu^2_{cr}}\right)^2\right]>0$,
we still have $\Psi_m\left(q=R_{cr}^2\right)<0$ in view of the
relation (\ref{rel}) and $\Psi_<\left(q=R_{cr}^2\right)=0$. But we
also have at the right end of the interval
$\Psi_m\left(q=q_*\right)<0$ and $\frac{d}{dq}\Psi_m|_{q_*}<0$,
demonstrating impossibility to infer positivity of the maximum of
$\Psi_m(q)$ inside $(R_{cr}^2,q_*)$ along these lines.

To circumvent this difficulty, we introduce an auxiliary function
$\Psi_a(\tilde{q},\tilde{\mu})$ of {\it two} real variables
$\tilde{q}$ and $\tilde{\mu}$ according to
\begin{equation}\label{psia} \Psi_a(\tilde{q},\tilde{\mu})=
2+\ln{\tilde{q}}-\tilde{q}
\tilde{\mu}^2-\left[2-\tilde{\mu}\right]^2\,.
\end{equation}
Note that $\Psi_m(q)$ in (\ref{psim}) is obtained from the above
function by replacing its arguments as $\tilde{q}\to q/R_{cr}^2$
and $\tilde{\mu}\to \frac{\mu(q)}{\mu_{cr}}$.

\begin{figure}
 \centering
 \includegraphics[width=10cm]{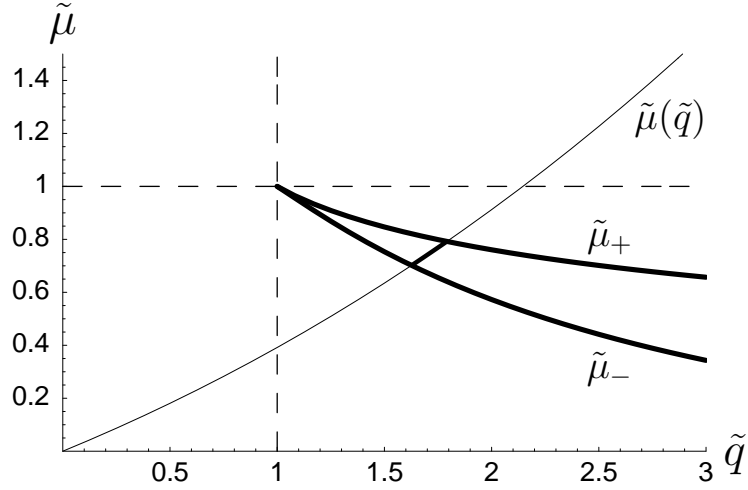}
 \caption{A representative function $\tilde{\mu}(\tilde{q})$ which is
 monotonically increasing and satisfies the condition $\tilde{q}_{*}>1$.
 These conditions are sufficient to ensure that the corresponding $\Psi_{m}(q)$ has a
 positive maximum leading to a positive complexity of minima.}\label{fig:2dplot}
\end{figure}

Considering values of $\Psi_a(\tilde{q},\tilde{\mu})$
 we easily find that the function can take positive values only in a wedge-like
 region of the $(\tilde{q},\tilde{\mu})$ plane
restricted by two boundary curves $\tilde{\mu}_{-}(q)$ and
$\tilde{\mu}_{+}(q)$, where
\begin{equation}\label{posi}
\tilde{\mu}_{\pm}(\tilde{q})=\frac{2\pm
\sqrt{D(\tilde{q})}}{1+\tilde{q}},\quad
D(\tilde{q})=4+(1+\tilde{q})\left(\ln{(\tilde{q})}-2\right)\,.
\end{equation}
and one has to require $D(\tilde{q})\ge 0$ for the existence of
these curves. Since $D(\tilde{q}=1)= 0$ and
$\frac{dD(\tilde{q})}{d\tilde{q}}=-1+\ln{q}+\frac{1}{q}\ge 0,
\,\forall q>0$, we conclude that $D(\tilde{q}\ge 1)\ge 0$. This
implies the region in between those two curves exists for
$\tilde{q}>1$, and for $\tilde{q}\to 1$ the two boundary curves
approach each other and meet at the point
$\tilde{\mu}_{\pm}(\tilde{q}=1)=1$ (see figure \ref{fig:2dplot}).

Now, the problem of finding the complexity of minima,
Eq.(\ref{complm}), for any monotonically increasing confinement
function $\mu(q)$ is equivalent to searching a maximum of the
function $\Psi_a(\tilde{q},\tilde{\mu})$ along the corresponding
curve $\tilde{\mu}(\tilde{q})=\mu{(q)}/\mu_{cr}$ in the plane (see
the figure). The point $q_*$ appearing in our analysis is nothing
else but the point of intersection of that curve with the horizontal
line $\tilde{\mu}=1$, and the condition $q_*>R_{cr}^2$ just means
that such an intersection happens at some value $\tilde{q}_*>1$. But
every such curve $\tilde{\mu}(\tilde{q})$ necessarily intersects the
region between  the boundary curves $\tilde{\mu}_{\pm}(\tilde{q})$,
and therefore necessarily has a portion along which the values of
$\Psi_a(\tilde{q},\tilde{\mu})>0$. This proves that the complexity
of minima is always positive as long as the condition $q_*>R^2_{cr}$
holds. This is equivalent to $\mu(R^2_{cr})<\mu_{cr}$ which is
precisely the replica symmetry breaking condition Eq.(\ref{zero1}).

Let us finally demonstrate that for a generic smooth confining
potential the complexity of minima vanishes always cubically at
criticality, the type of critical behaviour we have already found
in the simplest case of parabolic confinement. For this we follow
the same steps as in the end of the preceding section. Using the
parameter $\delta=1-\tilde{\mu}(1)\ll 1$ which controls the
distance to the criticality, we approximate in the critical region
$\tilde{q}=1+\epsilon,\,\epsilon\ll 1$, and use for the function
$\tilde{\mu}(\tilde{q})$ its Taylor expansion. Truncating as
before at first order: $\tilde{\mu}(\tilde{q})\approx
1-\delta+\epsilon \tilde{\mu}'(1)$ and substituting this for
$\tilde{\mu}$ in (\ref{psia}), we obtain
\begin{equation}\label{psiaa}
\Psi_{m}\left(\tilde{q}=1+\epsilon\right)=-2\delta^2
+\left[2\delta+4\tilde{\mu}'(1)\delta-\delta^2)\right]\epsilon-
\left[\frac{1}{2}+2\tilde{\mu}'(1)+2(\tilde{\mu}'(1))^2+O(\delta)\right]
\epsilon^2+O(\epsilon^3)\,.
\end{equation}
To find the corresponding complexity, we should maximize this over
$\epsilon$. To the leading order in $\delta$ the maximum is
attained at $\epsilon_0=2\delta/\left(1+2\tilde{\mu}'(1)\right)$.
Substituting this value back to (\ref{psiaa}) we immediately find
that all terms of the order of $\delta^2$ cancel, and thus at
criticality the complexity of minima behaves at least as
$\Sigma_s= C\,\delta^3$. To prove that the coefficient $C\ne 0$
and to find its actual value in terms of
$\tilde{\mu}'(1),\tilde{\mu}''(1)$ requires going to the terms of
order $\epsilon^2$ in the Taylor expansion for
$\tilde{\mu}(\tilde{q})$, and to the order $\epsilon^3$ in
(\ref{psiaa}). This is a straightforward but boring exercise, and
the resulting expression is not very elegant apart from the
simplest case of the parabolic confinement when
$\tilde{\mu}'(1)=\tilde{\mu}''(1)=0$, and $C=1/3$ as we have
already found before.

\subsection{On anomalous critical behavior of
complexities for hard-wall confining potentials.}

A fairly universal type of the critical behavior of both
complexities $\Sigma_s$ and $\Sigma_m$ revealed in the preceding
sections should be compared with the results reported recently in
our short communication \cite {FSW}, and independently in
\cite{BrayDean}. In both papers the landscape was confined in a
finite box, chosen in \cite{FSW} to be spherical of extent
$|\x|\le L=R\sqrt{N},\quad 0<R<\infty$. On top of such a hard-wall
confinement a harmonic confining potential was superimposed to
allow a comparison with the results of statistical mechanics of
the same problem studied by the replica trick in \cite{FS}. For
the sake of clarity, we will consider below only pure hard-wall
confinement, when the complexities for a given value of the radius
$R$ were found to be given by \cite{BrayDean,FSW}:
\begin{equation}\label{complolda}
\Sigma_{s}\left(R\right)=\ln{\left(\frac{R}{R_{cr}}\right)}\quad
\mbox{for}\,\, R\ge R_{cr},
\,\mbox{and}\,\Sigma_{s}\left(R\right)\le 0\,\,
\mbox{otherwise}\,,
\end{equation}
and
\begin{equation}\label{comploldb}
\Sigma_{m}\left(R\right)=-1+\ln{\left(\frac{R}{R_{cr}}\right)}\quad
\mbox{for}\,\, R\ge R_m=eR_{cr},
\,\mbox{and}\,\Sigma_{m}\left(R\right)\le 0\,\,
\mbox{otherwise}\,,
\end{equation}
with $R_{cr}$ given by Eq.(\ref{9}).
 Taking into account that the domain of
zero-temperature replica symmetry breaking in this particular
model is just given by $R>R_{cr}$\cite{FS}, we see that apparently
the type of behavior exemplified by Eqs.
(\ref{complolda}),(\ref{comploldb}) is very different from what we
have discussed earlier in this paper.  Namely, although in this
case the cumulative complexity $\Sigma_s$ is also positive
everywhere in the phase with broken replica symmetry for
$R>R_{cr}$, the complexity of minima $\Sigma_s$ becomes positive
only starting from a larger confining radius $R_m=eR_{cr}>R_{cr}$.
In other words, for the interval $R_{cr}<R<R_m$ the broken
ergodicity is not at all accompanied by the exponentially many
minima in the energy landscape. Another peculiarity is that the
complexity $\Sigma_s$ vanishes linearly rather than quadratically
with the distance $\delta_R=R/R_c-1$ close to the ergodicity
threshold.

In the framework of the present approach the function $\mu(q)$
describing the hard-wall confinement with a radius $R$  is
formally $\mu(q)=0$ for $q<R^2$ and $\mu(q)=\infty$ for $q\ge
R^2$. Such a form is apparently highly singular, and does not
immediately fit into the analysis of the preceding sections. To
circumvent this difficulty and to understand better the origin of
the hard-wall behavior (\ref{complolda}),(\ref{comploldb}) within
a more general framework we consider a family of non-smooth
confining potentials of the form:
\begin{equation}\label{family}
\mu(q)=\left\{\begin{array}{c}0\,,\quad \mbox{for}\,\, q<R^2\\
\,\mu\sqrt{h\left(\frac{q}{R^2}\right)}\,,\quad \mbox{for}\,\,
q\ge R^2
\end{array}\right.
\end{equation}
where $\mu>0$ is a control parameter, and $h(x)$ is an increasing,
non-negative smooth concave function $h(x)> 0,h'(x)\ge 0,h''(x)\ge
0$ for $\forall x>1$. We also assume $h(1)=0$ to ensure the
continuity of $\mu(q)$, and thus existence of a single solution
$q=q_*$ of the equation $\mu(q)=\mu_{cr}$ which as we know plays
an important role in our analysis. We will be in particular
interested in understanding the behavior of complexities
$\Sigma_s(R)$ and $\Sigma_m(R)$ in the limit $\mu\gg 1$, where the
confinement described by Eq.(\ref{family}) should approach a
hard-wall form. We also assume the confinement radius to satisfy
$R>R_{cr}$, to allow for a positive complexity.

The cumulative complexity of stationary points for this class of
potentials is given according to (\ref{complsfin}) by
$\Sigma_s^{(\mu)}(R)=\frac{1}{2}\mbox{max}_{\tilde{q}\in(1,q_*)}
\Psi_s^{(\tilde{\mu})}(\tilde{R},\tilde{q})$ where
\begin{equation}\label{psinsmth}
\Psi_s^{(\tilde{\mu})}(\tilde{R},\tilde{q})=\left\{\begin{array}{l}\ln{\tilde{q}}\,,\qquad\qquad\qquad
\mbox{for}\,\,
0<\tilde{q}<\tilde{R}^2\\\ln{\tilde{q}}+\tilde{\mu}^2h(\tilde{q}/\tilde{R}^2)(1-\tilde{q})\,,\quad
\mbox{for}\,\, \tilde{R}^2\le \tilde{q}\le \tilde{q}_*
\end{array}\right.
\end{equation}
and we used as usual the scaled parameter
$\tilde{\mu}=\mu/\mu_{cr}$ and the scaled confinement radius
$\tilde{R}=R/R_{cr}>1$ as well as $\tilde{q}=q/R_{cr}^2$, with the
value of the parameter $\tilde{q}_*$ being fixed by the condition
$\tilde{\mu}^2h(\tilde{q_*}/\tilde{R}^2)=1$. Differentiating the
expression (\ref{psinsmth}) over $\tilde{q}$ gives:
\begin{equation}\label{dir1}
\frac{d\Psi_s}{d\tilde{q}}|_{\tilde{q}\rightarrow
\tilde{R}^2+}=\frac{1}{\tilde{R}^2}-
\tilde{\mu}^2\left[h'(1)(\tilde{R}^2-1)\right]\,,
\end{equation}
which is clearly negative for large enough $\tilde{\mu}$, except
for the special case $h(1)=h'(1)=0$. On the other hand,
$\frac{d\Psi_s}{d\tilde{q}}|_{0<\tilde{q}<\tilde{R}^2}=\frac{1}{\tilde{R}^2}>0$.
We conclude that the function
$\Psi_s^{(\tilde{\mu})}(\tilde{R},\tilde{q})$ for large enough
$\mu$ has its local maximum precisely at $\tilde{q}=\tilde{R}^2$.
Actually, this is the global maximum, as
\begin{equation}\label{dir2}
\frac{d^2\Psi_s}{d\tilde{q}^2}|_{\tilde{q}>\tilde{R}^2}=-\frac{1}{\tilde{R}^4}\left\{1+
\tilde{\mu}^2\left[2\tilde{R}^2\,h'(\tilde{q}/\tilde{R}^2)+h''(\tilde{q}/\tilde{R}^2)(\tilde{q}-1)\right]\right\}<
0\,.
\end{equation}
so that
$\frac{d\Psi_s}{d\tilde{q}}<\frac{d\Psi_s}{d\tilde{q}}|_{\tilde{q}\rightarrow
\tilde{R}^2+}< 0$ for all $\tilde{R}^2<\tilde{q}<\tilde{q}_*$. We
immediately see that the cumulative complexity is then given
simply by $\frac{1}{2}\ln{\tilde{R}^2}$.

Let us now shortly discuss the complexity of minima along the same
lines. The analogue of the expression (\ref{psinsmth}) follows
from (\ref{psim}), and is given by
\begin{equation}\label{psinmmth}
\Psi_m^{(\tilde{\mu})}(\tilde{R},\tilde{q})=\left\{\begin{array}{l}\ln{\tilde{q}}-2\,,\qquad\qquad\qquad
\mbox{for}\,\,
0<\tilde{q}<\tilde{R}^2\\\ln{\tilde{q}}-2+4\tilde{\mu}\sqrt{h(\tilde{q}/\tilde{R}^2)}
-\tilde{\mu}^2(1+\tilde{q})h(\tilde{q}/\tilde{R}^2)\,,\quad
\mbox{for}\,\, \tilde{R}^2\le \tilde{q}\le \tilde{q}_*
\end{array}\right.
\end{equation}
Differentiating the above expression, we again easily see that
$\frac{d\Psi_m}{d\tilde{q}}|_{\tilde{q}\rightarrow
\tilde{R}^2+}<0$ for $\mu\gg 1$, except for the case
$h(1)=h'(1)=0$. Since
$\frac{d\Psi_m}{d\tilde{q}}|_{\tilde{q}\rightarrow \tilde{R}^2-}$
is always positive, the function $\Psi_m(\tilde{R},\tilde{q})$ has
its (global) maximum at $\tilde{q}=\tilde{R}^2$ and the complexity
is indeed given by the expression (\ref{comploldb}).

To investigate the behavior of both complexities in the remaining
exceptional case $h(1)=h'(1)=0$, we consider a particular example
$h(x)=(x-1)^2$. This implies
$\tilde{q}_*=\tilde{R}^2\left(1+\frac{1}{\tilde{\mu}}\right)$ and
for the case of cumulative complexity we are seeking to maximize
\begin{equation}\label{dir3}
\Psi_s(\tilde{q})=\ln{\tilde{q}}+\tilde{\mu}^2(\tilde{q}/\tilde{R}^2-1)^2(1-\tilde{q})
\end{equation}
over $\tilde{q}$ in the interval $ \tilde{R}^2\le \tilde{q}\le
\tilde{q}_*$. For large $\tilde{\mu}$ it is actually more
convenient to introduce a new variable $y\in[0,1]$ via
$\tilde{q}=\tilde{R}^2(1+y/\tilde{\mu})$ , and approximate
$\Psi_s(\tilde{q})$ with
\begin{equation}\label{dir33}
\Psi_s(y)=\ln{\tilde{R}^2}+y^2(1-\tilde{R}^2)+O(1/\tilde{\mu})\,,
\end{equation}
In view of the inequality $\tilde{R}^2>1$, the above expression
attains its maximum at $y=0$, and the cumulative complexity
$\Sigma_s(R)$ is again given by the same expression
(\ref{complolda}) as in all other cases of this family in the
hard-wall limit $\mu\gg 1$.

Curiously enough, the complexity of minima $\Sigma_m(R)$ in the
case $h(1)=h'(1)=0$ shows a behavior different from
(\ref{comploldb}). Taking as before $h(x)=(x-1)^2$, we need this
time to maximize
\begin{equation}\label{psina}
\Psi_m(\tilde{q})=\ln{\tilde{q}}-2+4\tilde{\mu}(\tilde{q}/\tilde{R}^2-1)
-\tilde{\mu}^2(1+\tilde{q})(\tilde{q}/\tilde{R}^2-1)^2\,
\end{equation}
over $\tilde{q}$ in the interval $ \tilde{R}^2\le \tilde{q}\le
\tilde{q}_*=\tilde{R}^2\left(1+\frac{1}{\tilde{\mu}}\right)$.
Introducing again the variable $y\in[0,1]$ via
$\tilde{q}=\tilde{R}^2(1+y/\tilde{\mu})$, and assuming $\mu\gg 1$
we replace (\ref{psina}) with
\begin{equation}\label{psinb}
\Psi_m(y)\approx\ln{\tilde{R}^2}-2+4y+O(1/\tilde{\mu})
-y^2(1+\tilde{R}^2)\,,
\end{equation}
which has its maximum at $y_0=2/(1+\tilde{R}^2)<1$. The
corresponding complexity is given by
\begin{equation}\label{psinc}
\Sigma_m=\ln{\tilde{R}^2}-2+\frac{4}{1+\tilde{R}^2}\,.
\end{equation}
which is indeed different from (\ref{comploldb}).
 In particular, the last term in (\ref{psinc}) ensures that
 the complexity of minima is positive everywhere in the phase with
 broken ergodicity $\tilde{R}>1$, and vanishes linearly when
 approaching the critical value $\tilde{R}>1$.

Thus, we have demonstrated that for every curve in the family
(\ref{family}) the cumulative complexity in the limit $\mu\to
\infty$ is indeed given by
$\Sigma_s(R)=\frac{1}{2}\ln{\tilde{R}^2}\equiv \ln{R/R_{cr}}$, in
full agreement with the hard-wall confinement formula
(\ref{complolda}). As to the complexity of minima, in the limit
$\mu\to \infty$ it is generically given by (\ref{comploldb}),
although the result may change if the hard-wall profile vanishes
smooth enough when approaching the point of non-analyticity,
$q=R^2$.

\section{Conclusions and open questions}
Let us briefly summarized our findings.
 We have demonstrated that for a generic, smooth
concave confining potentials with a continuous
 positive derivative the complexity extracted from
the mean number of totality of stationary points in the energy
landscape is positive simultaneously with the complexity
corresponding to the mean number of minima. The domain of
parameters where those complexities are positive is precisely one
where the zero-temperature limit of statistical mechanics in such
a landscape requires for its description the concept of broken
replica symmetry/broken ergodicity, Eq.(\ref{zero}).

On the other hand, for a non-analytic (hard-wall) confinement the
boundary of  non-ergodic behaviour coincides in general with the
domain of positive total complexity of saddle points but not
necessarily the minima. Moreover, the above analysis clearly
demonstrates that all peculiar features specific to the hard-wall
confinement are due to the discontinuity of the derivative of the
confining potential.

 In the case of a smooth confinement the complexity of
minima vanishes cubically when approaching the critical
confinement (\ref{zero1}), whereas the cumulative complexity
vanishes quadratically with the distance to criticality.  In the
appendix E we follow the method by Bray and Dean \cite{BrayDean}
and investigate the (annealed) complexities of stationary points
at a fixed value of the number of the negative eigenvalues of the
Hessian, the so-called {\it index} $\cal I$ of a critical point.
Restricting the consideration to the simplest case of parabolic
confinement Eq.(\ref{Hpar}) we reveal that the only stationary
points with non-vanishing complexity arising precisely at the
critical confinement $\mu=\mu_{cr}$ are those for which the number
of negative directions stays of order of unity in the large$-N$
limit: $\alpha=\lim_{N\to \infty}\frac{\cal I}{N}=0$. As we move
inside the glassy phase away from the critical value
$\mu=\mu_{cr}$, stationary points with an increasing range of
indices start to have a positive complexity. This is precisely the
mechanism behind the change from cubic vanishing typical for the
complexity of minima to the quadratic vanishing found for the
total complexity. Whether the stationary points which appear
precisely at the critical point are strictly minima, or may
include admixture of saddle points with a few negative eigenvalues
remains an interesting question deserving further, more elaborate,
investigation.

In our opinion, the results of present research suggest a few more
immediate questions which are worth understanding. For example,
the problem of investigating the magnitude of sample-to-sample
fluctuations of the number of stationary points is clearly very
interesting, though technically challenging. Another natural
extension would be to consider our model at finite temperatures.
This project would require defining properly an analogue of the
{\it free} energy (TAP-like, \cite{TAP}) landscape for the present
model, and recovering the whole transition line to the phase with
broken ergodicity from counting the corresponding stationary
points. Some steps in this direction were taken in \cite{FSW}, but
the present level of understanding remains far from being
satisfactory.

Finally, let us mention that the model considered in the present
paper is very intimately related \cite{FS} to the much-studied
spherical model of spin glasses, which essentially corresponds to
a particular choice of the confining potential
$V_{con}(\x)=\delta(\x^2-N)$ which forces the vector $\x$ to span
the sphere of radius $\sqrt{N}$. It is natural to expect that this
rather singular limit can be approached in the present method by
e.g. considering the confinement of the form $U(z)=\mu(z-1)^2$ and
allowing for $\mu\to \infty$. This expectation is supported by
recent rigorous results \cite{BDG} demonstrating that the
dynamical equations known for the spherical spin glass model are
faithfully reproduced in this limiting procedure. Let us however
note that the above choice goes beyond the immediate scope of the
present paper as we decided to restrict our attention by
considering only monotonically increasing confining potentials.
Actually, investigating other classes is not at all a problem, and
is certainly worth doing in view of the mentioned correspondence
with the spherical model, where quite a few results on
complexities of extrema of free energy landscapes were already
reported in the literature, see for example a good review
\cite{spherical} and references therein. We relegate this issue,
as well as a few other possible generalisations to subsequent
publications.

\appendix

\vskip 1cm

{\noindent \large \bf APPENDICES}

\section{On a geometric origin of the identity (\ref{ident})}
  Let us start with recalling the well-known Poincare-Hopf
index theorem. For any vector field ${\bf F}$ over any compact
manifold $M$ with no zeroes on the boundary $\partial M$ holds the
identity $Ind({\bf F})+Ind(\partial_{-}{\bf F})=\chi(M)$, where
$\chi(M)$ is a topological invariant, the Euler characteristic of
the manifold. The index $Ind$ of a vector field ${\bf F}$ is the
sum of indices of all the singular points. i.e. sum of indices
$\mbox{sign} \det{(\partial_i{\bf F}_j)}|_{{\bf x}={\bf x}_k}$
corresponding to all $N_s$ isolated zeroes ${\bf x}_k$ of the
field: ${\bf F}({\bf x}_k)=0$ . The field $\partial_{-}{\bf F}$ is
a vector field defined in terms of the values of ${\bf F}$ on a
subset of the boundary $\partial M$, such that the original field
${\bf F}$ points inward on this subset of the boundary. Precise
definition of the boundary component of the formula is immaterial
for our purposes.

Consider a random surface $H(\x)=\frac{1}{2}{\bf x}^T \hat{A} {\bf
x}+V(\x)$, with ${\bf x}\in \mathbb{R}^N$, $V({\bf x})$ being a
mean-zero random Gaussian function with the covariance
Eq.(\ref{cov1}), and the first term being a quadratic form
involving an arbitrary non-singular real symmetric matrix
$\hat{A}$, with $\det{\hat{A}}\ne 0$. Take any compact manifold
$M$ in $\mathbb{R}^N$ such that (i) all the stationary points
${\bf x}_k$ of $H(x)$ are with probability one belong to the
interior of that manifold and (ii) along the boundary $\partial M$
of the manifold the influence of random potential is already
negligible (the simplest choice of $M$ would be a ball $|{\bf
x}|\le L$ of a very large radius $L\to \infty$). Then topological
properties of the gradient field ${\bf F}=\partial H({\bf
x})/\partial {\bf x}$ along the boundary $\partial M$ are
determined by the gradient of the first term, i.e. by the vector
field ${\bf F}_{A}=\hat{A}{\bf x}$, so that $Ind(\partial_{-}{\bf
F})=Ind(\partial_{-}{\bf F}_{A})$. Then the Poincare-Hopf theorem
implies $Ind(\partial H({\bf x})/\partial {\bf x}) +
Ind(\partial_{-}{\bf F_{A}})=\chi(M)$. On the other hand, the
field ${\bf F}_{A}$ has the only zero at the origin, so that
$Ind({\bf F}_A)\equiv\mbox{sign} \det \hat{A}$, and the
Poincare-Hopf theorem appied to ${\bf F}_A$ requires $\mbox{sign}
\det \hat{A} + Ind(\partial_{-}{\bf F}_A)=\chi(M)$. Comparing
these two relations we see that
\begin{equation}\label{chi}
Ind(\partial H({\bf x})/\partial {\bf x})\equiv
\sum_{k=1}^{N_s}\mbox{sign} \det{\left(\frac{\partial^2
H}{\partial{\bf x}\partial {\bf x}}\right)|_{{\bf x}={\bf x}_k}}=
\mbox{sign} \det \hat{A}\,
\end{equation}
for every realization of the random surface $H({\bf x})$.

 On the other hand, the Dirac's
 $\delta$-functional measure satisfies the fundamental identity
\begin{equation}\label{delta}
\delta\left(\frac{\partial H}{\partial {\bf
x}}\right)\,|\det\left(\frac{\partial^2 H}{\partial{\bf x}\partial
{\bf x}}\right)|=\sum_{k=1}^{N_s}\delta({\bf x}-{\bf x}_k),
\end{equation}
which in fact underlies the Kac-Rice formula Eq(\ref{KR}).  As
above, the summation over ${\bf x}_k$ goes over all isolated zeros
of the gradient $\partial H/\partial {\bf x}$, and
$\delta\left(\frac{\partial H}{\partial {\bf
x}}\right)\equiv\prod_{i=1}^N\delta\left(\frac{\partial
H}{\partial x_i}\right)$. This fact allows us to convert the sum
in the left hand-side of (\ref{chi}) to an integral and rewrite
(\ref{chi}) as the identity
\begin{equation}\label{chi1}
\int_{\mathbb{R}^N}\det\left(\frac{\partial^2 H}{\partial {\bf
x}\partial{\bf x}}\right)\delta\left(\frac{\partial H}{\partial
{\bf x}}\right)\,d{\bf x}= \mbox{sign} \det \hat{A}\,
\end{equation}
valid for any realization of the random potential $V({\bf x})$. We
therefore can average (\ref{chi1}) over the realizations, and take
into account (i) independence of the first and second derivatives
for Gaussian-distributed random functions, and (ii) stationarity
of the random potential, Eq.(\ref{cov1}). In this way we come to
the relation
\begin{equation}\label{vel}
\mbox{sign} \det
\hat{A}=\left\langle\det\left(\hat{A}+\frac{\partial^2 V}{\partial
{\bf x}\partial{\bf
x}}\right)\right\rangle\,\int_{\mathbb{R}^N}\prod_{i=1}^N\left\langle\delta\left(\frac{\partial
H} {\partial x_i}\right)\right\rangle \,d{\bf x}
\end{equation}
Using the standard Fourier integrals representation for the
$\delta-$functional factors, and performing the averaging over the
gaussian gradients with the covariances ( cf. (\ref{4}))
$\left\langle\frac{\partial V} {\partial x_i}\frac{\partial V}
{\partial x_j}\right\rangle=-\frac{1}{N}f'(0)\delta_{ij}$, we see
that
\begin{eqnarray}
\left\langle \prod_{i=1}^N \delta\left(\frac{\partial H} {\partial
x_i}\right)\right\rangle &=& \int e^{-i{\bf u}^T \hat{A}{\bf
x}-\frac{1}{2N}{\bf u}^T{\bf u}|f'(0)|}\frac{d{\bf u}}{(2\pi)^N}\\
&=&\frac{1}{(2\pi|f'(0)|/N)^{N/2}}e^{-\frac{N}{2|f'(0)|}{\bf x}^T
\hat{A}^T\hat{A}{\bf x}}
\end{eqnarray}
Substituting the last expression to Eq.(\ref{vel}) and performing
the Gaussian integral over $\mathbb{R}^N$ yields the factor
$\left[\det{\hat{A}^T\hat{A}}\right]^{-1/2}=1/|\det{\hat{A}}|$.
Then eq.(\ref{vel}) is reduced to the relation
\begin{equation}\label{identa}
\det \hat{A}=\left\langle\det\left(\hat{A}+\frac{\partial^2
V}{\partial {\bf x}\partial{\bf x}}\right)\right\rangle
\end{equation}
equivalent to the identity Eq.(\ref{ident}), which is thus
verified for non-singular real symmetric matrices $\hat{A}$. By
analytic continuation it is extended to singular case as well.

\section{Distribution of the diagonal element of the GOE resolvent}
Our goal is to calculate the probability distribution ${\cal
P}(G)$ of the diagonal element of the resolvent (\ref{reso}) for
GOE matrices $\hat{H}$, in the large-N limit, for a given real
value of $s$.

Following the standard route we first evaluate the characteristic
function
\begin{equation}\label{char}
\chi(p)=\langle \exp\{ip \,G_H(\x)\}\rangle _{GOE}
\end{equation}
Our first observation is that the result can depend on $\x$ only
via the modulus $|\x|$ due to the rotational invariance of the GOE
probability density. This implies that we can choose $\x=|\x|{\bf
e}$, with ${\bf e}=(1,0,\ldots,0)$, so that
\begin{equation}\label{reso1}
G_H(\x)={\bf x}^2\,{\bf e}^T\frac{1}{\hat{H}+s\hat{I}}\,{\bf
e}={\bf x}^2\sum_{n=1}^N\frac{({\bf e ,\bf
e_n})^2}{s+\lambda_n}\,,
\end{equation}
where $\lambda_n,\bf e_n$ stand for the eigenvalues and the
corresponding eigenvectors of the GOE matrix $ \hat{H}$, and
$({\bf e} ,{\bf e}_n)$ is the scalar product. Remembering that the
eigenvectors $\bf e_n$ and the eigenvalues $\lambda_n$ of the GOE
matrices are statistically independent, we perform the averaging
over the eigenvectors first. To this end we recall that $N$ GOE
eigenvectors are (i) mutually orthogonal and (ii) uniformly
distributed over the unit sphere $({\bf e}_n,{\bf e}_n)=1$. As is
well-known, these conditions imply that in the large-$N$ limit the
projections $v_n=({\bf e}, {\bf e}_n)$ behave like independent
Gaussian variables with zero mean and variance $\langle
v_n^2\rangle_{GOE}=1/N$. Denoting $\langle \ldots \rangle_{v}$ the
averaging over these variables, we can write for the
characteristic function
\begin{eqnarray}
&&\langle \exp\{ip \,G_H(\x)\}\rangle
_{v}=\prod_{n=1}^N\int_{-\infty}^{\infty}\frac{dv_n}{\sqrt{2\pi/N}}\exp\{-\frac{N}{2}v_n^2+ip\,{\bf
x}^2\frac{v_n^2}{s+\lambda_n}\}\\ \nonumber &=&\prod_{n=1}^N
\left[\frac{N}{N-2ip\,{\bf
x}^2\frac{1}{s+\lambda_n}}\right]^{1/2}=\prod_{n=1}^N\frac{[s+\lambda_n]^{1/2}}
{[s-2i\frac{p}{N}\,{\bf x}^2+\lambda_n]^{1/2}}\,.
\end{eqnarray}
Remembering the remaining averaging over the JPD of GOE
eigenvalues $\lambda_n$, we see that the characteristic function
(\ref{char}) in the large-$N$ limit can be written as the
expectation value of the ratio of square roots of the
characteristic polynomials
\begin{equation}\label{char1}
\chi(p)|_{N\to\infty}=\lim_{N\to \infty}
\left\langle\frac{\left[\det(s+\hat{H})\right]^{1/2}}
{\left[\det(s-2i\frac{p}{N}\x^2+\hat{H})\right]^{1/2}}\right\rangle_{GOE}
\end{equation}
Precisely that expectation value was already calculated earlier in
a different context, see Eq.(35) and Eq.(48) of the paper
\cite{EF}, with the result
\begin{equation}\label{char2}
\chi(p)|_{N\to\infty}=\exp\left\{\x^2\left[\frac{i}{2}s\,p-|p|\pi\nu_{sc}(s)\right]\right\},\quad
\nu_{sc}(s)=\frac{1}{2\pi}\sqrt{4-s^2},\,\,|s|<2\,.
\end{equation}
The distribution ${\cal P}(G)$ immediately follows from this
expression after the Fourier-transform. Defining
$\tilde{G}=G_H(\x)/\x^2$ we see that the quantity is
Cauchy-distributed:
\begin{equation}\label{Gdistr}
{\cal
P}(\tilde{G})=\left\{\begin{array}{c}\frac{\nu_{sc}(s)}{\pi^2\nu^2_{sc}(s)+
\left(\tilde{G}-\frac{s}{2}\right)^2},\quad |s|<2\\
\delta\left(\tilde{G}-\frac{s}{2}\right),\quad |s|>2
\end{array}\right.\,.
\end{equation}

\section{An overview of the Dean-Majumdar functional integral approach}

Our aim in this section is the calculation of large-$N$
asymptotics of the required GOE averages.  The most economic way
of arriving to the desirable expressions known to us relies on a
heuristic method introduced by Dean and Majumdar \cite{DeanMaj}.
We however have every reason to believe that one can arrive to the
same results by employing a rigorous (and, necessarily, tedious)
mathematical procedures described for a very closely related
problem in a paper by A. Boutet de Monvel, L.Pastur and M.
Scherbina \cite{BPS}.

The quantities of interest for us are the GOE averages featuring
in equations (\ref{denfac1}) and (\ref{minfac1}), i.e.:
\begin{eqnarray}\label{mul}
 \mathcal{D}_{s}(s)  &=&\left\langle \left| \det(\hat{H}+s\hat{I}) \right|
 \right\rangle_{\mathrm{GOE}}\nonumber\\
& \propto&
\left[\prod_{i=1}^{N}\int_{-\infty}^{\infty}d\lambda_{i}\right]
 \prod_{i=1}^{N}|\lambda_{i}+s|\prod_{1\leq i < j \leq N} |\lambda_{i}-\lambda_{j}
 |e^{ -\frac{N}{4} \sum_{i=1}^{N}\lambda_{i}^{2}},
 \\
 \mathcal{D}_{m}(s) &=&\left\langle \theta(\hat{H} + s \hat{I}) \det( \hat{H} + s\hat{I}) \right\rangle_{\mathrm{GOE}}
 \nonumber\\&\propto& \left[\prod_{i=1}^{N}\int_{-s}^{\infty} d\lambda_{i}\right]
 \prod _{i=1}^{N}(\lambda_{i}+s)\prod_{1\leq i<j\leq N} |\lambda_{i}-\lambda_{j}
 |e^{ -\frac{N}{4} \sum_{i=1}^{N}\lambda_{i}^{2}}.,
\end{eqnarray}
where $\theta$-factor is equal to unity when its argument is a
positive definite matrix and is zero otherwise, and $\lambda_i$
stands for $N$ real eigenvalues of the matrix $\hat{H}$. We have
used that in each case the functions to be averaged depend only on
the matrix eigenvalues. The average is taken over an ensemble of
matrices with the probability density $\mathcal{P}(\hat{H})
d\hat{H} \propto \exp(-N\mathrm{Tr} \hat{H}^2/4)$ invariant with
respect to orthogonal transformations $\hat{H}\to
O\hat{H}\hat{O}^T$, and this allows us to perform in a standard
way\cite{Mehta} the integration over $N(N-1)/2$ angular variables
(eigenvectors of $\hat{H}$). The procedure yields just an overall
normalization factor, and the remaining expression is given by
(\ref{mul}). At this point it is worth mentioning, that the GOE
averages of the type as above can be viewed as performed over a
Gibbs measure describing an interacting gas of eigenvalues, and
there are methods to developed a rigorous mean-field description
of such a model in the large-$N$ limit \cite{BPS}.

In a similar in spirit, but much less formal way Dean and Majumdar
suggested to calculate integrals of this type by replacing the
multiple integration over the eigenvalues $\lambda_{i}$ with a
functional integration over the {\it density} of eigenvalues,
defined as
\begin{equation}\label{denfun}
 \rho_{N}(\lambda) := \frac{1}{N}\sum_{i=1}^{N}\delta(\lambda - \lambda_{i}).
\end{equation}
Exploiting this density, we have the formal identities
\begin{eqnarray}\label{id1}
\prod_{i=1}^N\phi(\lambda_i)&=&\exp{N\int
\ln{\phi(\lambda)}\,\rho_N(\lambda)\,d\lambda}\,,\\
\prod_{i<j}^N\psi(\lambda_i,\lambda_j)&=&\exp{\frac{N^2}{2}\int
\ln{\psi(\lambda,\lambda')}\,\rho_N(\lambda)\rho_N(\lambda')\,d\lambda\,d\lambda'}\,\,,
\end{eqnarray}
for a suitable choice of functions $\phi(\lambda)$ and
$\psi(\lambda,\lambda')=\psi(\lambda',\lambda)$.

 In this way the integrands of the multivariable
 integrals (\ref{mul}) can be viewed as functionals $f[\rho]$ of the density $\rho_{N}(\lambda)$,
 and introducing the (functional) Dirac $\delta-$function,
 one can formally write for such type of integrals
\begin{equation}
 \left[\prod_{i=1}^{N}\int_{a}^{b} d\lambda_{i}\right] f[\rho_N] =
\int\mathcal{D}[\rho]f[\rho]\left[\prod_{i=1}^{N}\int_{a}^{b}
d\lambda_{i}\right]\delta[\rho - \rho_{N}]
\end{equation}
where $f[\rho]$ is the functional $f$ expressed in terms of the
density $\rho$.  The integrals over the eigenvalues on the right
hand side of this expression can be conveniently evaluated after
using the standard formal Fourier representation for the
functional $\delta-$ function:
\begin{eqnarray}
 \delta[\rho - \rho_{N}] &=& \int\mathcal{D}[\omega]\exp\left(i\,N \int_{-\infty}^{\infty}
  \omega(t)[\rho(t) - \rho_{N}(t)]dt\right),\\
 &=& \int\mathcal{D}[\omega]\exp\left( i\,N \int_{\-\infty}^{\infty} \omega(t)\rho(t)dt -i\sum_{i=1}^{N}\omega(\lambda_{i}) \right),
\end{eqnarray}
with a suitably normalized measure $\mathcal{D}[\omega]$.
 The integration over
the eigenvalues is now trivially performed, resulting in
\begin{equation}
 \left[\prod_{i=1}^{N}\int_{a}^{b} d\lambda_{i}\right] \delta[\rho - \rho_{N}] \propto
\int \mathcal{D}[\omega'] \exp \left( iN\int_{-\infty}^{\infty}
\omega'(t) \rho(t)dt + N \ln\left(
\int_{a}^{b}e^{-i\omega'(t)}dt\right)\right)
\end{equation}
For $N \gg 1$ the main contribution to the above integral should
come from the stationary point of the exponent with respect to
variations in the field $\omega'$ satisfying the equation
\begin{equation}
 \rho(\lambda) =  \begin{cases}\frac{\exp(-i\omega'(\lambda))}
{\int_{a}^{b} \exp(-i\omega'(\lambda'))d\lambda'} \qquad &\text{if
$a\leq \lambda \leq b$,}\\ 0 &\text{otherwise.} \end{cases}
\end{equation}
It is clear from the above that $\int_{a}^{b}\rho(\lambda)d\lambda
=1$, and that $-i\omega'(\lambda)=\ln{\rho(\lambda)}+const$ for
$a\leq \lambda \leq b$. This yields, up to an overall constant
factor, the relation
\begin{equation}
\left[\prod_{i=1}^{N}\int_{a}^{b} d\lambda_{i}\right] \delta[\rho
- \rho_{N}] \propto \exp \left( -N\int_{a}^{b}
\rho(t)\ln(\rho(t))dt \right)\,,
\end{equation}
Note, that looking at our system as a kind of "eigenvalue gas",
this factor is simply a standard entropic contribution associated
with the density of particles.

Let us apply this method for our problem, i.e. to the calculation
of the required GOE averages. Denoting the range of integration in
each case by $\mathcal{R}_{s}$ and $\mathcal{R}_{m}$, we have for
$j\in \{s,m\}$
\begin{equation}\label{funcavg}
 \mathcal{D}_{j}(s) = \int\mathcal{D}[\rho] \exp\left( -\frac{N^{2}}{2}\mathcal{G}_{j}[\rho] +
 N\mathcal{T}_{j}[\rho]\right),
\end{equation}
where
\begin{eqnarray}
 \mathcal{G}_{j}[\rho] &=& \frac{1}{2} \int_{\mathcal{R}_{j}}
  \lambda^{2}\rho(\lambda) d\lambda - \int_{\mathcal{R}_{j}}\int_{\mathcal{R}_{j}}
  \rho(\lambda)\rho(\lambda')\ln(|\lambda-\lambda'|) d\lambda d\lambda',\label{G}\\
\mathcal{T}_{j}[\rho] &=&
\int_{\mathcal{R}_{j}}\rho(\lambda)\ln(|\lambda + s|) d\lambda -
\int_{\mathcal{R}_{j}} \rho(\lambda)\ln\rho(\lambda) d
\lambda.\label{T}
\end{eqnarray}

In the limit $N \to \infty$ in (\ref{funcavg}) the main
contribution to the functional integral in (\ref{funcavg}) comes
obviously from the value of $\rho$ which minimises the functional
$\mathcal{G}_{j}$. The stationary condition is found in the
standard variational procedure after incorporating the
normalisation condition $\int_{\mathcal{R}_{j} } \rho(\lambda)
d\lambda =1 $ via a Lagrange multiplier.  The resulting integral
equation reads:
\begin{equation}
\frac{\lambda^{2}}{4} + C = \int_{\mathcal{R}_{j} }
\rho(\lambda')\ln(|\lambda - \lambda'|) d\lambda' \qquad \text{for
$\lambda \in \mathcal{R}_{j}$ }.
\end{equation}
Differentiating this equation with respect to $\lambda$ gives
\begin{equation}\label{inteq}
\frac{\lambda}{2} = \int_{\mathcal{R}_{j}}
\frac{\rho(\lambda')d\lambda'}{\lambda - \lambda'} \qquad
\text{for $ \lambda \in \mathcal{R}_{j}$},
\end{equation}
where the integral must now be understood as a Cauchy principal
value.

 Solution of singular
integral equations of the type (\ref{inteq}) is discussed
extensively in \cite{mus92}.  The necessary inversion formula
essentially depends on whether the solution is required to be
bounded at each end. In general, given a function $g(\lambda)$ for
$\lambda \in (a,b)$ and the integral equation
$$g(\lambda) = \int_{a}^{b}\frac{f(\lambda')d\lambda'}{\lambda-\lambda'}$$
there is a unique solution $f(\lambda)$ which remains bounded at
the endpoints $a$ and $b$ provided that holds
$$ \int_{a}^{b}\frac{g(\lambda')d\lambda'}{\sqrt{(b-\lambda')(\lambda'-a)}} = 0\,.$$
In our case, the above condition implies
\begin{eqnarray}
 0 &=& \int_{a}^{b} \frac{\lambda'd\lambda'}{\sqrt{(b-\lambda')(\lambda'-a)}}= \frac{\pi(a+b)}{2}
\end{eqnarray}
Hence, such a solution only exists in the case $a=-b\equiv L/2$,
and in the latter case is given by the inversion formula
\begin{eqnarray}
 \rho(\lambda) &=& \frac{\sqrt{L^2/4-\lambda^2}}{2\pi^2} \int_{-L/2}^{L/2}
 \frac{\lambda'd\lambda'}{\sqrt{L^2/4-\lambda'^2}(\lambda'-\lambda)},\\
 &=& \frac{\sqrt{L^2/4-\lambda^2}}{2\pi^2}\left[ \int_{-L/2}^{L/2} \frac{d\lambda'}{\sqrt{L^2/4-\lambda'^2}} +\lambda \int_{-L/2}^{L/2} \frac{d\lambda'}{\sqrt{L^2/4-\lambda'^2}(\lambda'-\lambda)} \right],\\
 &=& \frac{\sqrt{L^2/4-\lambda^2}}{2\pi},
\end{eqnarray}
where in the final line we have used the identity
\begin{equation}
 \int_{a}^{b}\frac{dx}{\sqrt{(b-x)(x-a)}(x-z)} = 0 \qquad \text{for $z \in (a,b)$ }.
\end{equation}
The normalization condition fixes $L$, and the resulting
eigenvalue density is given by
\begin{equation}\label{semi}
 \rho_{sc}(\lambda) = \frac{1}{2\pi}\sqrt{4-\lambda^{2}}, \qquad \text{for $\lambda \in [-2,2]$}
\end{equation}
and zero otherwise, which is just the well-known Wigner
semi-circle law. In such a case the only dependence on the
variable $s$ in the exponent of ${\cal D}_s(s)$ comes from the
first term $ \int_{-2}^{2}
\rho_{sc}(\lambda)\ln(|\lambda+s|)d\lambda$ in (\ref{T}) which
leads to Eq.(37) as described in the text.

At the same time, when evaluating ${\cal D}_m(s)$ we have an
additional constraint on the density $\rho$, as the latter must
vanish for $\lambda<-s$. Obviously, the solution (\ref{semi}) can
satisfy such a constraint only as long as $s>2$, and has to be
modified in the opposite case $s>2$. As shown in \cite{mus92}
there always exists a solution of this type of integral equations
which is bounded only at one end of the integration range. The
solution which remains bounded at the upper end of the integration
range is given by
\begin{equation}
 \rho_{DM}(\lambda) = \frac{1}{2\pi^{2}}\sqrt{\frac{L-s-\lambda}
 {\lambda+s}}\int_{-s}^{L-s}\sqrt{\frac{\lambda'+s}{L-s-\lambda'}} \frac{\lambda'd\lambda'}{\lambda-\lambda'},
\end{equation}
where $L$ is a constant to be determined by the normalization. The
integral was further evaluated by Dean and Majumdar and the
resulting density is given, for our choice of the GOE measure, by
\begin{equation}\label{DM}
 \rho_{DM}(\lambda) = \begin{cases} \frac{1}{4\pi} \sqrt{\frac{L -\lambda - s}{\lambda + s}}[L+2\lambda] & \text{if $0 \leq \lambda + s \leq L$,}\\
0 &\text{otherwise,}\end{cases}
\end{equation}
where this time
\begin{equation}
L = \frac{2}{3}\left(s+\sqrt{s^{2}+12}\right).
\end{equation}
Note that $s\to 2$ implies $L\to 4$ and the Dean-Majumdar density
(\ref{DM}) reverts in this limit to the Wigner semicircular law
(\ref{semi}).

 Inserting the above function $\rho_{DM}$ in the definitions
of $\mathcal{G}_{m}$ and $\mathcal{T}_m$, see (\ref{G}),(\ref{T})
and performing the integrations with help of \textsc{Mathematica}
yields
\begin{multline}
 \mathcal{G}_{m}[\rho_{DM}]\equiv G_{\min}(s)=\frac{1}{216}\left(72s^2-s^4-30s\sqrt{12+s^2}-s^3\sqrt{12+s^2}+54(3+4\ln 6)
 \right)\\
-\ln(s+\sqrt{s^2+12})\,,
\end{multline} which is our expression Eq.(69), as well as a formula for
$\mathcal{T}_m[\rho_{DM}]$. The latter formula is however
irrelevant for finding the complexity of minima, as our analysis
reveals that the complexity is determined by value $s\to 2$ when
obviously $\mathcal{T}_m[\rho_{DM}]\to\mathcal{T}_m[\rho_{sc}]$.

\section{Analysis of the replica-symmetric solution
for a spherically-symmetric confining potential.}

The calculation of the de-Almeida-Thouless condition in the
general energy surface of the form (\ref{Hgen}) requires only very
minor modifications in comparison with the case of a parabolic
confinement (\ref{Hpar}) considered in detail in \cite{FS}.
 Applying the procedure of Ref.\cite{FS}
yields the following exact expression for the averaged replicated
partition function:
\begin{equation}\label{replica2}
\left\langle Z_{\beta}^n\right\rangle={\cal C}_{N,n} N^{Nn/2}
e^{\frac{\beta^2}{2}Nnf(0)}\int_{Q>0}
\left(\mbox{det}Q\right)^{-(n+1)/2} e^{-\beta N\Phi_n (Q) }\, dQ
\end{equation}
where
\begin{equation}\label{repham1}
 \Phi_n (Q)=\sum_{a=1}^n U\left(\frac{q_{aa}}{2}\right)-
 \frac{1}{2\beta}\ln{(\det{Q})}-\beta\sum_{a<b}
f\left[\frac{1}{2}(q_{aa}+q_{bb})-q_{ab}\right]
\end{equation}
where $C_{N,n}$ is a known numerical constant and $N$ is assumed
to satisfy the constraint $N>n$.

The form of the integrand in Eq.(\ref{replica2}) is precisely one
required for the possibility of evaluating the replicated
partition function in the limit $N\to \infty$ by the Laplace
("saddle-point") method. The free energy is then given by
\begin{equation}\label{repfreeen}
F_{\infty}=\lim_{N\to \infty}\frac{1}{N}\langle
F\rangle=-\frac{T}{2}\ln(2\pi e)-\frac{1}{2T}f(0)+\lim_{n\to
0}\frac{1}{n}\Phi_n (Q)
\end{equation}
where the entries of the matrix $Q$ are chosen to satisfy the
stationarity conditions: $\frac{\partial \Phi_n(Q)}{\partial
q_{ab}}=0$ for $a\le b$. This yields, in general, the system of
$n(n+1)/2$ equations:
\begin{equation}\label{sp1}
\frac{1}{2}\mu\left(q_{aa}\right)-\frac{1}{\beta}\left[Q^{-1}\right]_{aa}-\beta\sum_{b(\ne
a)}^n f'\left[\frac{1}{2}(q_{aa}+q_{bb})-q_{ab}\right]=0,\quad
a=1,2,\ldots,n
\end{equation}
and
\begin{equation}\label{sp2}
 -\frac{1}{\beta}\left[Q^{-1}\right]_{ab}+\beta
f'\left[\frac{1}{2}(q_{aa}+q_{bb})-q_{ab}\right]=0,\quad a\ne b
\end{equation}
where $f'(x)$ stands for the derivative $df/dx$ and we used the
convention (\ref{not}).

The Replica Symmetric Ansatz amounts to searching for a solution
to Eqs.(\ref{sp1},\ref{sp2}) within subspace of matrices
$Q=Q_{RS}>0$ such that $q_{aa}=q_d$, for any $a=1,\ldots n$, and
$q_{a<b}=q_0$. The system of equations is easy to solve and to
obtain in the replica limit $n\to 0$ the following relations:
\begin{equation}\label{qsym}
q_d=\frac{T}{\mu_d}-\frac{1}{\mu_d^2}f'\left(\frac{T}{\mu_d}\right),\quad
q_0=-\frac{1}{\mu_d^2}f'\left(\frac{T}{\mu_d}\right)
\end{equation}
where we denoted for brevity $\mu_d\equiv \mu(q_d)$.

The stability analysis of this solution amounts to expanding the
function $\Phi_n(Q)$ in Eq.(\ref{repham1}) around the extremum
point up $Q=Q_{RS}$ to the second order in deviations:
$\Phi=\Phi_{SP}+ \delta \Phi + \frac{1}{2}{\delta}^2 \Phi$. The
stationarity condition amounts to $\delta \Phi =0$ yielding the
system (\ref{sp1})-(\ref{sp2}). The term ${\delta}^2 \Phi$ is a
quadratic form in independent fluctuation variables $\delta
q_{ab}, a\le b$ and can be generally written as ${\delta}^2
\Phi=\sum_{(ab),(cd)}\delta q_{(ab)} G_{(ab),(cd)}\,\delta q
_{(cd)}$. As usual the stable extremum corresponds to the positive
definite quadratic form, and along the critical line the quadratic
form becomes semi-definite. Checking positive definiteness of
${\delta}^2 \Phi$ amounts to finding the (generalized) eigenvalues
$\Lambda$ of the matrix
$G_{(ab),(cd)}=\frac{\partial^2\Phi}{\partial q_{ab}\partial
q_{cd}}$. It is easy to see that in our case the quadratic form
can be written as
\begin{equation}
{\delta}^2 \Phi=\frac{\mu'\left(q_d\right)}{2}\sum_{a}^n(\delta
q_{ab})^2+\frac{T}{2}\mbox{Tr}\left[\delta Q (Q_{RS})^{-1}\delta Q
(Q_{RS})^{-1}\right]-\frac{1}{T}\sum_{ab}f''(D_{ab})\delta
D_{ab}^2\,,
\end{equation}
where we introduced short-hand notations $\delta
D_{ab}=\frac{1}{2}(\delta q_{aa}+\delta q_{bb}-2\delta q_{ab})$
and $D_{ab}=q_d-q_{ab}$. The corresponding eigen-equations for
generalized $n(n+1)/2$ component eigenvectors $\eta_{(ab)}$ with
$a<b$ can be written straightforwardly by repeating the analysis
of \cite{FS} and are given for $T>0$ by:
\begin{equation}\label{unified}
 \sum_{cd} \left ( Q_{RS}^{-1} \right )_{ac}
 \eta_{(cd)} \left ( Q_{RS}^{-1} \right )_{db} + \frac{1}{T^2}
 f''(D_{ab}) (\delta
D_{ab})+\delta_{ab}\left[\frac{2}{T}\mu'\left(q_d\right)-
 \sum_{c}f''(D_{ac}) (\delta D_{ac})\right] = \Lambda^* \eta_{(ab)}
\end{equation}
 As the entries of the matrix $Q_{RS}$ are given by
$q_{ab}=q_0+(q_d-q_0)\delta_{ab}$, its inverse $Q^{-1}$ has the
same form $(Q^{-1})_{ab}=p_0+(p_d-p_0)\delta_{ab}$, with $p_0$ and
$p_d-p_0$ are given in the limit $n\to 0$ by
\begin{equation}\label{invsym}
p_0=-\frac{q_0}{(q_d-q_0)^2}=\frac{1}{T^2}f'\left(\frac{T}{\mu_d}\right),\quad
p_d-p_0=\frac{1}{q_d-q_0}=\frac{\mu_d}{T}
\end{equation}
Now we can follow faithfully the lines of the classical work by De
Almeida and Thouless \cite{AT} and to provide an explicit
construction of the families of eigenvectors with components
$\eta_{(ab)}$ of different symmetry. For the model with parabolic
confinement this construction was discussed in \cite{FS} and goes
through here without any modification. There are three different
families of eigenvectors,  first two yielding only eigenvalues
$\lambda^*$ with positive real part, hence stable. The dangerous
third family of eigenvectors is that satisfying the constraints:
\begin{equation}\label{cond2}
 \eta^{(III)}_{(aa)}=0,\,\forall\, a;,\quad \sum_{d}\eta^{(III)}_{(ad)}=0, \forall\,a
\end{equation}
The equations Eq.(\ref{unified}) are then reduced to a single
equation
\begin{equation}\label{ATeig}
\left[(p_d-p_0)^2-\frac{1}{T^2}f''(q_d-q_0)\right]\eta^{(III)}_{(ab)}=
\Lambda^*\eta^{(III)}_{(ab)},\quad \forall \,a\ne b.
\end{equation}
and substituting here (\ref{invsym}) we find that the replica
symmetric solution is stable as long as
\begin{equation}\label{ATcond}
\mu_d^2\ge f''\left(\frac{T}{\mu_d}\right)
\end{equation}
with the equality $\mu_d^2 = f''\left(\frac{T}{\mu_d}\right)$
providing the condition of the replica symmetry breaking
transition. Solving the latter together with Eq.(\ref{qsym}) is
easy after introducing the auxilliary variable
$\tau=\frac{T}{\mu_d}$. This gives finally the transition
temperature line in the form
\begin{equation}
T_{AT}=\tau\sqrt{f''(\tau)},\,
\end{equation}
where $\tau$ satisfies the equation
\begin{equation}\label{AT}
\quad
\mu\left(\tau-\frac{f'(\tau)}{f''(\tau)}\right)=\sqrt{f''(\tau)}\,.
\end{equation}

 In particular, the transition at zero temperature $T\to 0$
requires $\tau\to 0$, hence the replica symmetric solution at zero
temperature is stable as long as the inequality
\begin{equation}\label{zero}
\mu\left(-\frac{f'(0)}{f''(0)}\right)\ge \sqrt{f''(0)}\,.
\end{equation}
is satisfied, with the equality sign standing for the transition
condition to the region with broken replica symmetry.

\section{Complexity of stationary points with a given index}
In this appendix we outline the calculation of complexity for the
stationary points with a given index ${\cal I}(\hat{H})$ (the
number of negative eigenvalues of the Hessian $\hat{H}$) for our
model. We closely follow the method developed by Bray and Dean in
\cite{BrayDean}, which allows one to perform the calculation for
the extensive values of the index scaled with $N$ as ${\cal
I}=\alpha N$, where $0\le\alpha<1$. For simplicity we consider
only the case of parabolic confinement, with constant
$\mu(z)=\mu$.

Repeating the same steps as in \cite{YFglass}, the expected value
of the number ${\cal N}(\mu,\alpha)$ of stationary points with a
given value of $\alpha$ can be straightforwardly shown to be given
by (cf. Eq.(\ref{JPD2})):
\begin{eqnarray} \label{E1}
&&\left\langle \mathcal{N}(\mu,\alpha)
\right\rangle=\frac{1}{s^{N}}
\frac{1}{2^N}\sqrt{\frac{2}{N+2}}\left(\frac{N}{\pi}
\right)^{N(N+1)/4}
\\ \nonumber &\times &\int d\hat{H} |\det(s I_{N} + \hat{H})|\delta\left( \mathcal{I}(sI_{N} +
\hat{H})-N\alpha \right)e^{- \frac{N}{4}\left[\Tr\hat{H}^{2}-
\frac{1}{N+2}(\Tr\H)^{2}\right]} .
\end{eqnarray}
where we denoted $s = \mu/\sqrt{f''(0)}\equiv\mu/\mu_{cr}$.

The integration over the real, symmetric matrix $\H$ is in the
standard way reduced to an integral over its eigenvalues
$\lambda_i$ , and further to the functional integral over the mean
eigenvalue density, Eq.(\ref{denfun}), see Appendix B. This gives:
\begin{equation}\label{nfrac}
 \left\langle \mathcal{N}(\mu,\alpha)\right\rangle = \frac{1}{s^{N}}\frac{\Psi(\mu,\alpha)}{D},
\end{equation}
where
\begin{eqnarray}
\nonumber \Psi(\mu,\alpha) & = & \int \mathcal{D}[\rho] \exp\left(
-N^{2}\mathcal{S}_{2}[\rho] + N\mathcal{S}_{1}[\rho,s]\right) \\
\label{psidef} & & \quad \times
\delta\left(\int_{-\infty}^{\infty} \rho(\lambda)d\lambda -1
\right) \delta\left(\int_{-\infty}^{-s}\rho(\lambda)d\lambda -
\alpha \right). \label{psii}
\end{eqnarray}
The expressions appearing in the exponential in (\ref{psidef}) are
given by
\begin{eqnarray}
\nonumber \mathcal{S}_{2}[\rho] &=&
\frac{1}{4}\int_{-\infty}^{\infty}\rho(\lambda)\lambda^{2}d\lambda
- \frac{1}{4}\left( \int_{-\infty}^{\infty}\rho(\lambda)\lambda
d\lambda\right)^{2} \\ & & \quad-
\frac{1}{2}\int_{-\infty}^{\infty}\int_{-\infty}^{\infty}
\rho(\lambda)\rho(\lambda')\ln(|\lambda-\lambda'|)d\lambda
d\lambda',
\end{eqnarray}
and
\begin{eqnarray}
\nonumber  \mathcal{S}_{1}[\rho,s] &=&
\int_{-\infty}^{\infty}\rho(\lambda)\ln(|s+\lambda|) d\lambda
-\int_{-\infty}^{\infty}\rho(\lambda)\ln\rho(\lambda)d\lambda \\
& & \quad -\frac{N}{2(N+2)}\left(
\int_{-\infty}^{\infty}\rho(\lambda)\lambda d\lambda\right)^{2}.
\end{eqnarray}
The denominator $D$ appearing in (\ref{nfrac}) is a normalization
factor which can also be represented as a functional integral and
is given by
\begin{eqnarray}\label{ddef}
 D = \int \mathcal{D}[\rho] \exp( - N^{2}\mathcal{S}_{2}[\rho] + N\mathcal{S}_{1}'[\rho] )
 \delta\left(\int_{-\infty}^{\infty}\rho(\lambda)d\lambda -1 \right),
\end{eqnarray}
where
\begin{equation}\label{s1prime}
 \mathcal{S}_{1}'[\rho] = -\int_{-\infty}^{\infty}\rho(\lambda)\ln\rho(\lambda)d\lambda -
 \frac{N}{2(N+2)}\left(\int_{-\infty}^{\infty}\rho(\lambda)\lambda d \lambda \right)^{2}.
\end{equation}
 As $N \to \infty$, the main contribution to the functional integral in (\ref{psidef})
comes from the function $\rho$ which minimizes
$\mathcal{S}_{2}[\rho]$.  By using the definition of the mean
eigenvalue,
\begin{equation}
 \bar{\lambda} := \int_{-\infty}^{\infty}\rho(\lambda)\lambda d\lambda,
\end{equation}
and further introducing the function $f$ defined as
\begin{equation}
 f(x) = \rho(x + \bar{\lambda}),
\end{equation}
we rewrite $\mathcal{S}_{2}[\rho]$ in the following form:
\begin{equation}\label{s2f}
 \mathcal{S}_{2}[f] = \frac{1}{4}\int_{-\infty}^{\infty}f(x)x^2dx -
 \frac{1}{2}\int_{-\infty}^{\infty}\int_{-\infty}^{\infty}f(x)f(x')\ln(|x-x'|)dx dx'.
\end{equation}
We now need to find the function $f$ which minimizes (\ref{s2f})
subject to the constraint $\int f(x)dx = 1$. A simple variational
calculation (cf. Appendix C) shows that the minimizer is the usual
Wigner semi-circle density, given by
\begin{equation}\label{fdist}
 f(x) = \frac{\sqrt{4-x^2}}{2\pi}.
\end{equation}
The value of $\bar{\lambda}$ can now be fixed using the
restriction on the index given by the second $\delta-$functional
factor in (\ref{psii}). This leads to
\begin{equation}\label{halpha}
 \alpha = \frac{2}{\pi}\int_{-1}^{-h/2} \sqrt{1-x^2}dx,
\end{equation}
where $h = s + \bar{\lambda}$.  Each value of $\alpha\in[0,1]$
then corresponds to a unique value of $h\in[0,2]$.

We can apply the same variational procedure to the normalization
factor $D$. The only difference is that the functional integration
in the denominator does not contain the restriction on index,
which results in second term in ${\cal S}_1'$ minimized by the
value $\overline{\lambda}=0$. On the other hand, the terms
containing $\mathcal{S}_{2}$ appear in both $D$ and $\Psi$, and
therefore cancel from their ratio in (\ref{nfrac}) when we apply
the asymptotic evaluation of the integrals in the limit $N\to
\infty$. Likewise, the term of the form
$\int\rho(\lambda)\log\rho(\lambda)d\lambda$ will cancel between
the numerator and the denominator. Consequently, the relation
(\ref{nfrac}) is reduced asymptotically to
\begin{eqnarray}\label{lll}
 \left\langle \mathcal{N}(\mu,\alpha)\right\rangle&\sim&\frac{1}{s^N}
 \exp\left(N\left[\int_{-2}^{2}f(x)\ln(|x+h|)-\frac{1}{2}
\bar{\lambda}^{2} \right]\right),
\end{eqnarray}
where $\overline{\lambda}=h-s$, and $h$ is related to $\alpha$ by
Eq.(\ref{halpha}). For the semicircular form of $f(x)$ the
integral above can be explicitly calculated:
\begin{equation}
 \frac{1}{2\pi}\int_{-2}^{2}\sqrt{4-x^2}\ln|x+h|\,dx = -\frac{1}{2} +
 \frac{h^{2}}{4}\,.
\end{equation}
Substituting this result to Eq.(\ref{lll}) yields the final form
of the complexity corresponding to a given index:
\begin{equation}\label{complexity}
 \Sigma(s,\alpha) = -\frac{1}{2} + \frac{h^2}{4} - \frac{(h-s)^{2}}{2}-\ln s.
\end{equation}

The analysis of this expression is convenient to perform
separately for $\alpha=0$ and $\alpha>0$.
\begin{enumerate}
 \item The relation (\ref{halpha}) for $\alpha = 0$ implies $h=2$.
 Inserting this in (\ref{complexity}) gives
\begin{equation}
 \Sigma(s,0) = \frac{1}{2} - \frac{(2-s)^{2}}{2} - \log s \,.
\end{equation}
This expression is zero at $s = 1$, and taking the derivative with
respect to $s$ gives
\begin{eqnarray}
 \frac{\partial}{\partial s} \Sigma(s,0)  =  2 - \left( s + \frac{1}{s}
 \right)\le 0
\end{eqnarray}
with equality achieved only for $s=1$. We see
 that $\Sigma(s,0)$ is decreasing with $s$ and hence is positive
for $0<s<1$. Also $\frac{\partial^2}{\partial s^2} \Sigma(s,0) =
\frac{1}{s^2}-1
 ,\quad
 \frac{\partial^3}{\partial s^3}\Sigma(s,0) =  -\frac{2}{s^3}$.
 Thus, the first non-vanishing derivative at $s=1$
is the third derivative, implying that the complexity of
stationary points with $\alpha=0$ must vanish cubically as $s \to
1$. This agrees with the general analysis of complexity of minima
performed earlier in the paper.

\item
Eq.(\ref{halpha}) implies $h<2$ for $\alpha>0$. Differentiating
(\ref{complexity}) with respect to $s$ gives
\begin{equation}
 \frac{\partial}{\partial s} \Sigma(s,\alpha) = h - \left(s + \frac{1}{s} \right)<0.
\end{equation}
Thus, $\Sigma(s,\alpha)$ is strictly decreasing as a function of
$s$, and vanish {\it linearly} at some critical point $s_{cr}$
which can be easily found numerically as a function of $\alpha$.
Observing that
\begin{equation}
 \Sigma(1,\alpha) = -\frac{(h-2)^2}{4}<0\,.
\end{equation}
we conclude that the critical value $s_{cr}(\alpha)$ must satisfy
$s_{cr}(\alpha)<1$ for any $\alpha>0$. In other words, the
complexity of stationary points with any extensive index ${\cal
I}=O(N)$ is still negative at the point of ergodicity breaking
$s=1$ (i.e. $\mu=\mu_{cr}=\sqrt{f''(0)}$ in the original
notations), and starts to be positive already inside the phase
with broken ergodicity: $\mu_{cr}(\alpha>0)<\sqrt{f''(0)}$.

\end{enumerate}

 If we considered the annealed average of the total number of stationary points, rather than
those with a fixed index, this would be equivalent to integrating
$\langle\mathcal{N}(\mu,\alpha)\rangle$ over all values of
$\alpha$. In the limit $N\gg 1$ such an integral will obviously be
dominated by the value of $\alpha$ which maximizes
$\Sigma(s,\alpha)$. As $d\alpha/dh\ne 0$ for $|h|<2$  the maximum
occurs when
\begin{eqnarray}
 0 = \frac{\partial}{\partial h} \Sigma(s,\alpha) = s -
 \frac{h}{2}\,,
\end{eqnarray}
so that $h=2s$. At this point of maximum
\begin{equation}
 \Sigma_{tot}(s) = \frac{s^2 - 1}{2} - \log s.
\end{equation}
This shows that complexity of all stationary points tends to zero
as $s\to 1$. and by taking derivatives we find that at $s=1$ the
first derivative vanishes but the second one is non-zero.
Consequently, the (annealed) complexity related to the total
number of stationary points vanishes as $(1-s)^2$ as $s \to 1$, as
indeed was found in \cite{YFglass} by a different method.

Moreover, the last result can be used to show that the critical
value $s_{cr}(\alpha)$ , - defined as $s$ at which the complexity
with a given $\alpha$ vanishes, - must monotonously decrease with
$\alpha$ increasing. First observe that in fact we have shown that
$\Sigma(h/2,\alpha)>0$ which implies $s_{cr}(\alpha)>h/2$. The
value $s_{cr}(h)$ by definition solves $\Sigma(s_{cr}(h),h)=0$. As
$h$ decreases with $\alpha$ increasing, it is enough to show that
$ds_{cr}/dh>0$. Taking the derivative with respect to $h$ and
rearranging gives
\begin{eqnarray}
 \frac{d s_{cr}}{d h} = -\left(\frac{\partial \Sigma /\partial h}{\partial \Sigma / \partial s} \right)
 =  \frac{s_{cr}-h/2}{s_{cr}+s_{cr}^{-1}-h}> 0.
\end{eqnarray}
as required.

\end{document}